\documentclass{aa}

\usepackage{graphicx}
\usepackage{txfonts}
\usepackage{adjustbox}
\usepackage{hyperref}
\usepackage{color}
\newcommand{\xmm}{\emph{XMM--Newton}~}

\begin{document}

   \title{Ultra-thick warm absorbers: \\Enlarging the parameter space of AGN ionised outflows}

%   \subtitle{Dissertation on a pretty little dog}

   \author{R. Middei
          \inst{1,2,3}\fnmsep\thanks{riccardo.middei@inaf.it}
          \and E. Piconcelli\inst{1}
           \and E. Nardini \inst{4}
           \and A. Marinucci \inst{5}
           \and M. Laurenti \inst{6,1}
            \and A. Luminari\inst{7,1} \and \\
            A. Trindade Falc\~ao\inst{8,3}
            \and A. Tortosa\inst{1}
           \and M. Perri \inst{1,2}
            \and S. Puccetti \inst{5}
            \and D. Kr\'ol \inst{3,9}
                \and L. Borrelli\inst{10,11}}

   \institute{
   INAF Osservatorio Astronomico di Roma, Via Frascati 33, 00078 Monte Porzio Catone (RM), Italy
   \and Space Science Data Center, Agenzia Spaziale Italiana, Via del Politecnico snc, 00133 Roma, Italy
    \and Center for Astrophysics | Harvard \& Smithsonian, 60 Garden Street, Cambridge MA 02138, USA
   \and INAF -- Osservatorio Astrofisico di Arcetri, Largo Enrico Fermi 5, I-50125 Firenze, Italy
\and ASI -Agenzia Spaziale Italiana, Via del Politecnico snc, 135
Roma, Italy
\and Dipartimento di Fisica, Università degli Studi di Roma Tor
Vergata, Via della Ricerca Scientifica 1, 135 Roma, Italy
\and INAF - Istituto di Astrofisica e Planetologia Spaziali, Via del Fosso del Cavaliere 100, 00133, Roma, Italy
\and NASA Goddard Space Flight Center, Code 662, Greenbelt, MD 20771, USA
\and Astronomical Observatory of the Jagiellonian University, Orla 171, 30-244 Krak\'ow, Poland
\and Dipartimento di Fisica e Astronomia Augusto Righi, Università di Bologna, via Gobetti 93/2, 40129 Bologna, Italy
 \and INAF – OAS, Osservatorio di Astrofisica e Scienza dello Spazio Bologna, Via Piero Gobetti, 93/3, 40129 Bologna, Italy       }

   \date{Received mm/dd/yyyy; accepted mm/dd/yyyy}

% \abstract{}{}{}{}{}
\abstract
  % Context (opzionale, ma consigliato)
   {The analysis of X-ray absorption features in active galactic nuclei (AGN) provides a wealth of information about the physical properties of the matter surrounding supermassive black holes (SMBHs). While standard correlations between the ionisation state, column density, and velocity typically distinguish between disc winds and warm absorbers, some sources exhibit properties that  significantly deviate from these trends.}
  % Aims
   {We investigate a class of X-ray absorbers, which we define as ultra-thick warm absorbers (UTWAs), identified in a sample of 12 AGN. These absorbers are characterised by exceptionally high column densities and ionisation parameters (($\log (N_{\rm H}/\rm cm^{-2})\gtrsim22.5$ and $0.5 \lesssim \log(\xi/\rm erg~cm~s^{-1}) \lesssim 2.5$)) that lie outside the typical ranges observed in standard warm absorbers.}
  % Methods
   {We performed detailed X-ray spectral analyses of both unpublished and archival {\it XMM-Newton}, {\it NuSTAR}, and {\it Swift} datasets to characterise the physical properties of UTWAs in four of these twelve sources. We studied their variability on timescales ranging from days to years and explored their connection with other spectral features.}
  % Results
   {All AGN hosting UTWAs in our sample exhibit extreme soft X-ray variability, in some cases up to an order of magnitude, primarily driven by changes in the absorbing gas. In a subset of these sources (four out of 12), the UTWAs are accompanied by signatures of ultra-fast outflows (UFOs) in the Fe K$\alpha$ energy range.}
  % Conclusions
   {UTWAs represent a rare but crucial phase of AGN feedback. We discuss their physical origin, their potential connection with UFOs, and provide insights into why these high-column density, unusually ionised absorbers appear so rarely in local AGN samples.}


   \maketitle
%
%-------------------------------------------------------------------
\section{Introduction}
%\begin{figure*}[]
 \begin{figure}
   \centering
   \includegraphics[width=\columnwidth]{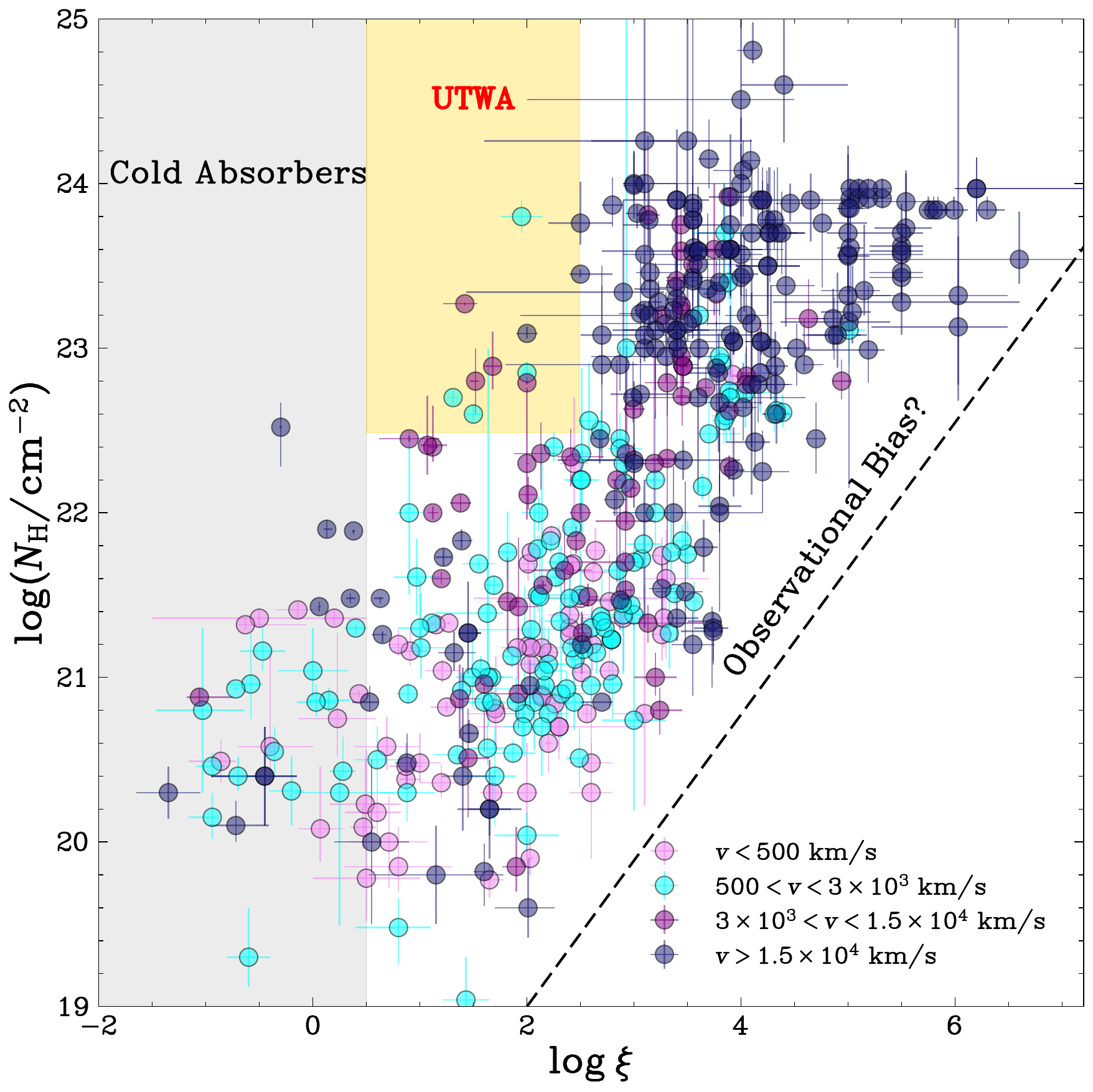}
  \caption{Distribution of the different absorbers observed in the X-ray spectra of AGN, as listed and classified by \citet[][see Section 1 for details]{Yamada2024}. We further divided the absorbers into bins of velocity ($v$\,$<$\,500 km s$^{-1}$,
    500\,$<$\,$v$\,$<$\,$3\times10^3$ km s$^{-1}$, $3\times10^3$\,$<$\,$v$\,$<$\,$1.5\times10^4$ km\ s$^{-1}$, $v$\,$>$\,$1.5\times10^4$ km s$^{-1}$). Faster absorbers appear to preferentially exhibit larger $\xi$ values. Detecting highly ionised absorbers with low column densities is  observationally challenging with current facilities, and the lower-right corner of the plane is basically unpopulated. Remarkably, another region remains scarcely occupied, i.e. the golden-shaded area which corresponds to absorbers with moderately high $\xi$ and very high $N_{\rm H}$, which we refer to as ultra-thick warm absorbers (UTWAs). }\label{yamada_2024}
   \end{figure}
\begin{table*}[!]
\caption{Log of the observations analysed in this work.}
    \centering

    \begin{tabular}{c@{\hspace{27pt}}c@{\hspace{18pt}}c@{\hspace{18pt}}c@{\hspace{18pt}}c@{\hspace{18pt}}c@{\hspace{18pt}}c}
        \hline
		Source & Detector & Obs. ID & Start date & Net exp. & rates &Obs.\\
        &&&(yyyy-mm-dd) & (ks) & (counts s$^{-1}$) &\\

		%& & &yyyy-mm-dd & ks &\\
		\hline
		SDSS\,J080243.40+310403.3 & pn/mos1/2&0862770801  &2021--04--29  & 44/53/53& 0.28/0.08/0.08&1\\
		&XRT &00014794001  &2021--09--05&  3& 0.07&2\\
		&XRT &00014794002  &2021--09--09  & 3& 0.09&3\\
   \hline
   		SDSS\,J080908.13+461925.6&pn/mos1/2&0843830101  &2019--09--04  & 12/16/16& 0.03/0.007/0.007& 1\\
		&pn/mos1/2&0843830201  &2022--04--04  & 51/58/59 &0.08/0.02/0.02 & 2 \\
   \hline
 PG\,1535+547 &pn &0150610301  &2002--11--03  & 29 &0.14 & 1\\
		  &pn &0300310301  &2006--01--16  & 27& 0.35&2\\
        &pn &0300310401  &2006--01--22  & 29&0.52 &3\\
		&pn &0300310501  &2006--01--24  & 25& 0.36&4\\
        &pn&0790590101  &2016--09--12  & 46&0.05 & 5\\
      	&FPMA/B&60201023002  &2016--09--12  & 58/58&0.01/0.01 &5\\
        &pn&0790590201  &2016--09--14  & 36&0.05 &6\\
        &FPMA/B&60201023004  &2016--09--14  & 82/82&0.01/0.01 &6\\
   \hline
 WISEA\,J023228.79+202349.9   &pn/mos2&0604210201  &2009--08--21 & 18&0.03/0.01 &1\\
        &pn/mos2 &0604210301  &2009--08--23  & 20& 0.04/0.02&2\\
		&mos2 &0810821801  &2021--08--08  & 120& 0.12&3\\
        &FPMA/B&10702609004  &2021--08--11& 58/57& 0.03/0.01&3\\
        &mos2&0902110201  &2023--01--15  & 15&0.06& 4\\
   \hline
\end{tabular}\label{log}
\\[3pt]
\small\textbf{Notes:} The reported net exposures correspond to the effective integration time after accounting for detector dead time and the removal of periods affected by high background flaring. Both net exposures and count rates have been rounded to the nearest significant digit for clarity.
\end{table*}

\indent Reprocessing of the primary X-ray continuum by intervening matter along our line of sight can significantly reshape the X-ray spectra of AGN. Of particular interest from an evolutionary perspective is the impact of the ionised, outflowing gas phases located within a few parsecs from the central engine \citep[e.g.,][]{Laha2021N}. Absorption signatures from moderately ionised (i.e. `warm') gas are observed in the X-ray spectra of at least 50\% of AGN  \citep[e.g.][]{Reynolds1997,George1998,Porquet2004,Piconcelli2005,Laha2014}. These warm absorbers (WAs hereafter) are typically found in outflow, although with modest velocities of $v_{\rm out}$ $\approx$\,100--500 km s$^{-1}$, and they are characterised by an ionisation parameter\footnote{The ionisation parameter, defined as $\xi=L_{\rm ion}/nr^2$ (erg cm s$^{-1}$), depends on the ratio between the ionising flux received by the gas and its density. For simplicity, its units are omitted throughout the text.} in the range $\log \xi$\,$\sim$\,0--2 and an equivalent hydrogen column density $N_{\rm H}$\,$\lesssim$\,10$^{22}$ cm$^{-2}$. In keeping with these properties, warm absorption mostly affects the soft X-ray band through the presence of species such as O\,\textsc{vii}--\textsc{viii}, Fe\,\textsc{xiv}--\textsc{xvi}, Ne\,\textsc{ix}--\textsc{x}, and Mg\,\textsc{xi}--\textsc{xii}, whose spectral footprints are conspicuous even at low spectral resolution \citep[e.g.][]{Nardini2014,Cappi2016,Middei2023}. At the same time, the reflection gratings on board {\it Chandra} and \xmm allowed for extensive, high-resolution studies of WAs in the brightest AGN well before the advent of X-ray micro-calorimeters \citep[e.g.][]{McKernan2007,Mehdipour2010,Reeves2013,Detmers2011,Mao2019}. \\
\indent The opposite end of the $N_{\rm H}$--$\xi$ parameter space with respect to WAs is populated by the so-called ultra-fast outflows (UFOs), which are probed via highly blueshifted Fe\,\textsc{xxv}--\textsc{xxvi} absorption lines in the $\sim$\,7--10 keV band, implying the existence of nuclear disc winds ejected with velocities faster than $v_{\rm out}$ $\sim$\,10,000 km s$^{-1}$ \citep[e.g.][]{Tombesi2010,Gofford2013,Matzeu2023,Gianolli2024}. Disc winds do have indeed large column densities (i.e. $N_{\rm H}$\,>\,10$^{23}$ cm$^{-2}$) and a high ionisation state (i.e. $\log \xi$\,>\,4). Ostensibly, the combination of velocity and transported mass  inferred from the X-ray data would result in substantial kinetic power \citep{Nardini2015,Tombesi2015}, capable of playing a major role in the AGN feedback process, which is expected to regulate both super massive black holes (SMBH) growth and host-galaxy evolution \citep[][]{Kormendy2013,King2015} by enhancing or quenching star formation activity \citep[e.g.][]{Faucher2012,Zubovas2013,Zubovas2014,Cresci2015,Feruglio2015,Bischetti2019,Marasco2020}.

Based on velocity and ionisation arguments, `standard' WAs are usually placed at much larger distances from the primary X-ray source with respect to UFOs \citep[e.g.][]{Blustin2005}, and their driving mechanism is likely different as well \citep{Mizumoto2019}. Moreover, WAs are expected to provide a negligible contribution to AGN feedback \citep[][]{Krongold2007,Laha2016}. On this wake, the tentative correlation between $\log N_{\rm H}$ and $\log\xi$ based on large compilations of X-ray absorbers in AGN \citep[e.g.][]{Tombesi2013} might partly be an observational artefact, as suggested by the evidence that the most highly ionised winds (typically UFOs with $\log\xi$\,$\gtrsim$\,4) are also the thickest ones: the identification of lower-N$_{\rm H}$, high-ionisation absorbers might be hindered by the limited collecting area of current X-ray detectors in the Fe-K band (see also \citealt{Crenshaw2012}).

The scenario gets even more complicated when also the velocity information is considered. UFOs with moderate ionisation ($\log \xi$\,$\lesssim$\,2) have indeed been detected in the soft X-rays in several sources \citep[e.g.][]{Gupta2015,Longinotti2015,Reeves2016}, with $v_{\rm out}$\,$\geq$\,10$^4$ km s$^{-1}$.

In addition,  high-ionisation Fe-K absorbers with  $v_{\rm out}$ $\sim$ a few $\times$10$^3$ km s$^{-1}$  have been also reported  \citep[][]{Risaliti2005b,Turner2008,Risaliti2011,Mehdipour2017}.
 The origin of these components is likely tied to the innermost circum-nuclear gas, as their velocities and ionisation states are consistent with highly ionised, clumpy material  crossing our line of sight at distances comparable to the Broad Line Region (BLR).
Remarkably, whenever high-quality data are available, multiple outflow components that cover a wide range of column density, ionisation, and/or velocity are revealed in the same object, strongly suggesting a physical connection between the different gas phases. For instance, widely different ionisation states are known to coexist at the same outflow velocity in several disc winds \citep[][]{Kriss2018,Reeves2020,Mehdipour2022}, while in other cases the lower ionisation components turn out to be somewhat slower, but still much faster than a typical WA \citep[e.g.][]{Krongold2021}. The emerging picture of multi-phase X-ray outflows, possibly involving stratification and/or clumpiness \citep[e.g.][]{Serafinelli2019}, has now been spectacularly confirmed by the first high-resolution UFO observations obtained with XRISM/Resolve \citep[][]{Xrism2025,Mehdipour2025,Xu2025,Xiang2025}.

In this context, however, the extent of any physical link between WAs and UFOs remains largely uncertain. Recently, \citet{Yamada2024} compiled a collection of the physical properties of 583 X-ray winds, consisting of WAs
and UFOs from the observations of 132 AGN, adopting the parameters reported in the literature. The latter work confirmed the previously established $\log N_{\rm H}$--$\log\xi$ correlation, although with a large scatter and possible observational biases at low $N_{\rm H}$ and high $\xi$. We show an adapted version of Fig.~4 by \citet{Yamada2024} in Figure \ref{yamada_2024}. Interestingly, different regions of the parameter space are underpopulated. When the absorber has very large $N_{\rm H}$ and low $\xi$ (corresponding to the upper-left corner of Figure \ref{yamada_2024}),
current X-ray observations cannot accurately constrain its ionisation level, making it indistinguishable from a cold/neutral absorber, such as those responsible for variable partial covering and X-ray eclipses \citep[e.g.][]{Nardini2011}.
Conversely, the region of the parameter space characterised by very high $\xi$ and low $N_{\rm H}$  is likewise inaccessible with current observations. As evident in Figure \ref{yamada_2024}, another region of the $\log N_{\rm H}$–$\log \xi$ plane remains scarcely populated. It corresponds to absorbers marked by high $N_{\rm H}$ and intermediate  $\xi$.  In this paper, we define the portion of the parameter space defined by $\log N_{\rm H}$\,$>$\,22.5 and 0.5\,$<$\,$\log \xi$\,$<$\,2.5 as the locus of ultra-thick warm absorbers (UTWAs), a term formally introduced in this work to distinguish them from standard WAs.
While these boundaries are somewhat arbitrary, they serve to identify the most extreme regime
of warm absorption, with column densities consistent with those of UFOs.
This selection allows us to focus on the `core' population of
these absorbers, thus providing a robust census against the typical $N_{\rm H} - \log \xi$
degeneracies of CCD-resolution spectra and potential selection biases. Even if this region should be plainly accessible to current detectors, only a handful of AGN exhibiting a UTWA have been found so far. Accordingly, this extreme class of WAs has been understandably overlooked. Yet, the study of UTWAs offers a promising opportunity to better understand such an unusual regime of ionised outflows, and it could be key to gaining deeper insights into the complex circumnuclear environment of SMBHs. As UTWAs share their ionisation state with the low-$\xi$ UFOs and their column density with the high-$\xi$ ones, some of them might even be related to the UFO phenomenon itself. In this sense, UTWAs could help reveal the full multi-phase structure of accretion disc winds.
In this work, we investigate the nature of UTWAs by analysing their variability behaviour. We present new and archival X-ray observations of four AGN (one of which falls well beyond the commonly explored $z$\,<\,0.1 range) that show the clear-cut presence of a UTWA along our line of sight in at least one epoch. Moreover, we report on additional objects from the literature, bringing the total number of AGN hosting UTWA to 12 (excluding confirmed UFOs that fall in the region of interest).

The paper is organised as follows: in Section 2 we report on the sample and on the data reduction procedure for the observations examined here for the first time, while in Section 3 we present the X-ray spectral analysis for each source and derive the physical parameters of the new UTWAs. In Sections 4 and 5 we discuss our results and draw our conclusions, respectively.

\begin{figure*}[th]
   \centering
   \includegraphics[width=0.85\columnwidth]{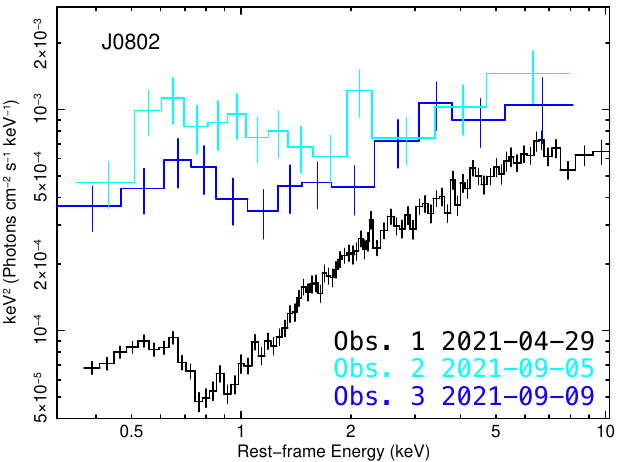}
    \includegraphics[width=0.85\columnwidth]{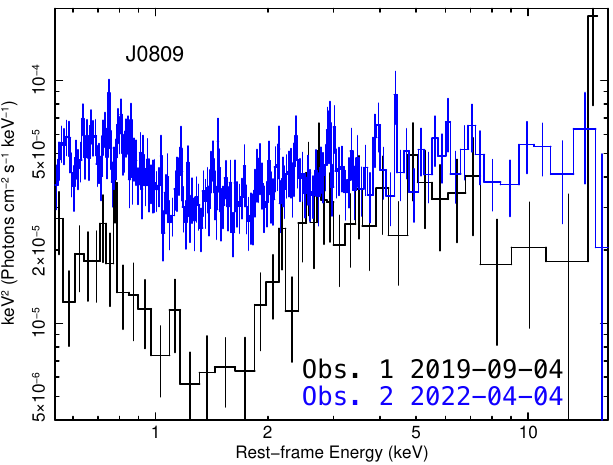}
    \includegraphics[width=0.85\columnwidth]{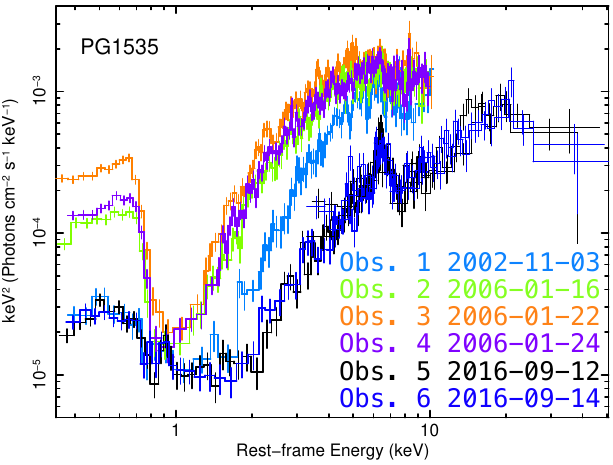}
    \includegraphics[width=0.85\columnwidth]{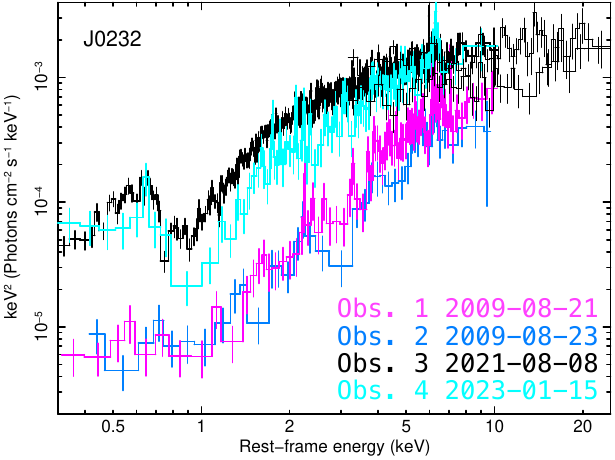}
     \caption{Unfolded X-ray spectra of the four AGN listed in Table~\ref{log}. Complex and variable absorption affects the soft X-ray band of all the observations presented here. For simplicity, only one spectrum is shown when more cameras where used.}
         \label{unfoldedall}
   \end{figure*}

\section{Sample and X-ray observations}
As mentioned above, only a small number of absorbers detected in the X-rays are compatible with being UTWAs, and their nature is still poorly understood. A few such absorbers are already reported in \citet[][]{Yamada2024}, as shown in Figure \ref{yamada_2024}. In this work we expand the available UTWA statistics by considering a sample of  AGN known to exhibit significant soft X-ray variability due to ionised absorption, excluding objects where transient obscuration events, lasting from hours to weeks/months, can be attributed to neutral or very lowly ionised gas clouds \citep[e.g.][]{Lamer2003}. In the following, we discuss in detail four distinct AGN in which our spectral analysis reveals the presence of UTWAs. In Appendix~\ref{Appendix}, we report a list of additional 8 AGN for which previous studies in the literature have provided evidence of UTWAs affecting their soft X-ray bands.

\subsection{New AGN hosting UTWAs}
All the sources analysed here are type-1 AGN exhibiting prominent absorption variability in the soft X-rays. These objects are listed below. \\
\textbf{SDSS\,J080243.40+310403.3} (referred to as J0802 hereafter) is a Seyfert 1 galaxy at a redshift of $z$\,=\,0.04, powered by an accreting SMBH with a mass of $\log (M_{\rm BH}/M_\odot)$\,$\sim$\,7.69  \citep[][]{Bennert2021} and a bolometric luminosity of $\log (L_{\rm bol}$/erg\,s$^{-1}$)\,$\sim$\,44 \citep[][]{Liu2019}. In this work, we analyse a $\sim$\,55 ks {\it XMM-Newton} observation taken in April 2021 and two snapshot observations of 3 ks each carried out by {\it Swift} on September 2021 (see Table 1 for further details). The X-ray spectral properties of this Seyfert galaxy are presented here for the first time.\\
\textbf{SDSS\,J080908.13+461925.6} (J0809 hereafter) is a quasar and is the most distant source in our sample, with a redshift of $z$\,=\,0.65537 \citep{Ahn2012}. For this source, \citet[][]{Laurenti2022} report the BH mass and bolometric luminosity to be $\log(M_{\rm BH}/M_{\odot})$\,$\approx$\,8.5 and $\log(L_{\rm bol}$/erg\,s$^{-1}$)\,$\sim$\,46.5, respectively. Its X-ray properties were first reported by the same authors who investigated a sample of highly accreting AGN at intermediate redshifts with {\it XMM-Newton}.  Based on a 18-ks observation carried out in 2019, these authors found that the X-ray spectrum is best fitted with a power-law model and measured  a 2--10 keV luminosity of 2.3\,$\times$\,10$^{44}$ erg s$^{-1}$.
They also mentioned the presence of a significant broad absorption trough in the soft X-rays of this source, but any in-depth analysis
was beyond the scope of that study. In this work, we present the complete analysis of this 2019 data set along with that of a deeper {\it XMM-Newton} observation taken in April 2022 (see Table 1). \\
\textbf{PG\,1535+547} is a nearby ($z$\,=\,0.038) and relatively bright Narrow Line Seyfert 1 galaxy \citep[][]{Constantin2003}, with $\log (L_{\rm bol}$/erg\,s$^{-1}$)\,$\sim$\,44.6, the most extensively studied source in our sample at X-rays.
Different estimates of the BH mass are available for this AGN. \citet{Vestergaard2006} and \citet{Zhang2006} derived $\log(M_{\rm BH}/M_{\odot})\,=\,$ 7.19 and 6.94, respectively, using virial relations based on the optical continuum luminosity and the FWHM of H$\beta$. Additionally, \citet{Zhang2006} provided a third estimate of $\log(M_{\rm BH}/M_{\odot})\,=\, 7.34$ by rescaling the stellar velocity dispersion inferred from the [O\,\textsc{iii}] line width. The first exposure of this AGN was presented in \citet{Schartel2005} and subsequently re-analysed by \citet{Ballo2008} together with three new \textit{XMM-Newton} observations (Obs.\,1--4 in Table 1) that caught the source in different flux states. These authors reported a very complex X-ray spectral shape with multiple, variable absorbers that heavily affect  the primary continuum. In this work, we re-analyse these early observations and the data sets obtained in September 2016 from a joint \textit{XMM-Newton} and \textit{NuSTAR} campaign (Obs.\,5--6), recently presented by \citealt{MadathilPG}\footnote{The focus of that paper, which appeared when our analysis had been already finalized, is different and largely complementary to ours, as the authors explore the origin of the broadband X-ray emission. Interestingly, also their models require the presence of a UTWA.}.\\
\textbf{WISEA\,J023228.79+202349.9} (J0232 hereafter) is classified as a broad-line AGN at $z$\,=\,0.029 \citep{Lansbury2017}. The X-ray properties of this source were first presented as part of the serendipitous source catalogue observed by \textit{NuSTAR} (see  \citealt{Zappacosta2018}). By analysing the broadband X-ray spectrum using {\it XMM-Newton} and \textit{NuSTAR} data, they found a very flat spectral shape likely due to the presence of a two-layer absorber, consisting of a cold layer and a partially ionised one. J0232 is located at about $\sim$\,8~arcmin from the well-known $\gamma$-ray blazar 1ES\,0229+200 \citep[$z$\,=\,0.14; e.g.][]{Ehlert2023}. Consequently, it has been serendipitously observed multiple times with \textit{XMM-Newton}. In some cases, however, it falls outside the field of view of at least one EPIC camera (Table~\ref{log}). The source was also serendipitously observed with \textit{NuSTAR} in coordination with \textit{XMM-Newton} (Obs.\,3 in Table~\ref{log}), but unfortunately also in this case J0232 lies at the very edge of the FPMB module\footnote{This affects the quality of the corresponding FPMB spectrum, which contains $\sim$\,50\% fewer counts compared to FPMA.}. For J0232, \citet{Lansbury2017} report BH mass and luminosity values of $\log(M_{\rm BH}/M_{\odot})$\,$\sim$\,7.8 and $\log (L_{\rm bol}$/erg\,s$^{-1}$)\,$\sim$\,45.2.

\subsection{Data reduction}

In the present study, we use X-ray data obtained from \textit{XMM-Newton} \citep[][]{Jansen2001}, \textit{NuSTAR} \citep[][]{Harrison2013}, and \textit{Swift} \citep[][]{Gehrels2004}. For each source, the available data sets are listed in Table~\ref{log}.

\indent We extracted the \textit{XMM-Newton} EPIC spectra \citep[pn and MOS;][respectively]{Struder2001,Turner2001} using the Science Analysis Software (\textsc{sas}, v.\,20211130) and the most recent calibration files (\textsc{ccf}, April 2024). The event files were filtered to remove any background flaring events ($>\,$0.4 counts s$^{-1}$).  %To extract
For the source spectrum, we adopted circular extraction regions with variable radii. The final radius was selected to maximise the signal-to-noise (S/N) ratio \citep[see][for details]{Piconcelli2004}. This procedure was applied to each observation and each detector. The background was estimated using a circular region (with radius of 70 arcsec) centred on a blank area of the sky. The spectra were then binned to ensure at least 5 counts per bin for J0802 and J0809, 25 counts per bin for J0232, and 30 counts per bin for PG\,1535+547. We note that the individual {\it EPIC} spectra, when available for the same observation, were found to be consistent with each other in both flux and spectral shape, within the uncertainties.

{\it NuSTAR} data are available only for PG\,1535+547 and J0232, which were both observed in the context of joint campaigns with \textit{XMM-Newton}. To extract the spectra from the two focal plane modules of \textit{NuSTAR} (FPMA/B), we used the \textit{NuSTAR} data analysis software (\textsc{NuSTARDAS}; see Perri et al.~2013),\footnote{\url{https://heasarc.gsfc.nasa.gov/docs/nustar/analysis/nustar_swguide.pdf}} adopting the latest calibration databases (CALDB v.\,20240325). For both modules we used circular extraction regions in the range of 20--50 arcsec, again with a variable radius selected to maximise the S/N of the source. The background was inferred using a region of the same size, located in a blank area of the same chip. We verified that using larger background areas does not affect the final spectral results. The spectra were then binned to achieve at least 25 counts per bin for J0232 and 30 counts per bin for PG\,1535+547.

\textit{Swift/XRT} \citep[][]{Burrows2005} observations were processed using the online tool from the multi-mission archive,\footnote{Available at the Space Science Data Center (SSDC) webpage, \url{https://www.ssdc.asi.it/mma.html}} which relies on the \textsc{XRTDAS} software\footnote{A complete guide is accessible at this link: \url{https://swift.gsfc.nasa.gov/analysis/xrt_swguide_v1_2.pdf}} and the latest available calibration databases. We utilised this on-the-fly tool to extract the source spectrum,  adopting a circular region with a radius of 20 pixels centred on the target (where 1 pixel corresponds to 2.36 arcsec for the {\it XRT} detector). An annulus with inner and outer radii of
40 and 80 pixels was used to extract the corresponding X-ray background spectrum. The source spectra were then binned using the standard command \texttt{grppha} to ensure a minimum of 3 counts per bin. \\
\indent In Figure \ref{unfoldedall}, we show the corresponding spectra that capture the full extent of the  variability observed in the sources analysed in this work. A simple visual inspection suggests that most of the spectral variations are apparently connected with changes in the opacity of the partially ionised absorbing matter.

\section{Spectral analysis}

\indent The spectral analysis presented here was performed using the fitting software package \textsc{XSPEC} \citep[][]{Arnaud1996}. All the spectra were fitted simultaneously for each AGN, and the statistics adopted, either $C$ \citep[after][]{Cash1979} or $\chi^2$ minimisation, was set by the lower quality spectrum available for that specific source. The Galactic column density, derived following \citet{HI4PI2016}, is always included via the \texttt{tbabs} model. We also allow for a cross-calibration constant ($K$) between the spectra obtained from different detectors in the same epoch.
The analysis was performed assuming the \textsc{XSPEC} built-in cosmology with $\Omega_{m}$\,$\geq$\,$1-\Lambda_0$ and the default values of $H_0$\,=\,70, $q_0$\,=\,0.0, and $\Lambda_0$\,=\,0.73. All the uncertainties reported in the tables correspond to a 90\% confidence level. We note that for the modelling of the UTWAs we adopted the \textsc{XSTAR}-based grid \texttt{zxipcf} \citep[][]{Miller2006,Reeves2008}, which has been widely used in previous studies and thus represents a well-established reference framework, allowing for a straightforward comparison with the results of \citet[][]{Yamada2024} and other literature works. Moreover, we intentionally adopt a phenomenological description of the continuum to maintain consistency across the different signal-to-noise ratios of our sample. Using the highest quality spectra (e.g. PG~1535+547), we verified that the inferred UTWA parameters are compatible when more complex modelling is applied (e.g. blurred reflection or warm Comptonisation), also in agreement with the findings of \citet{MadathilPG}.

\subsection{J0802}
\begin{figure}
	\centering
	\includegraphics[width=\columnwidth]{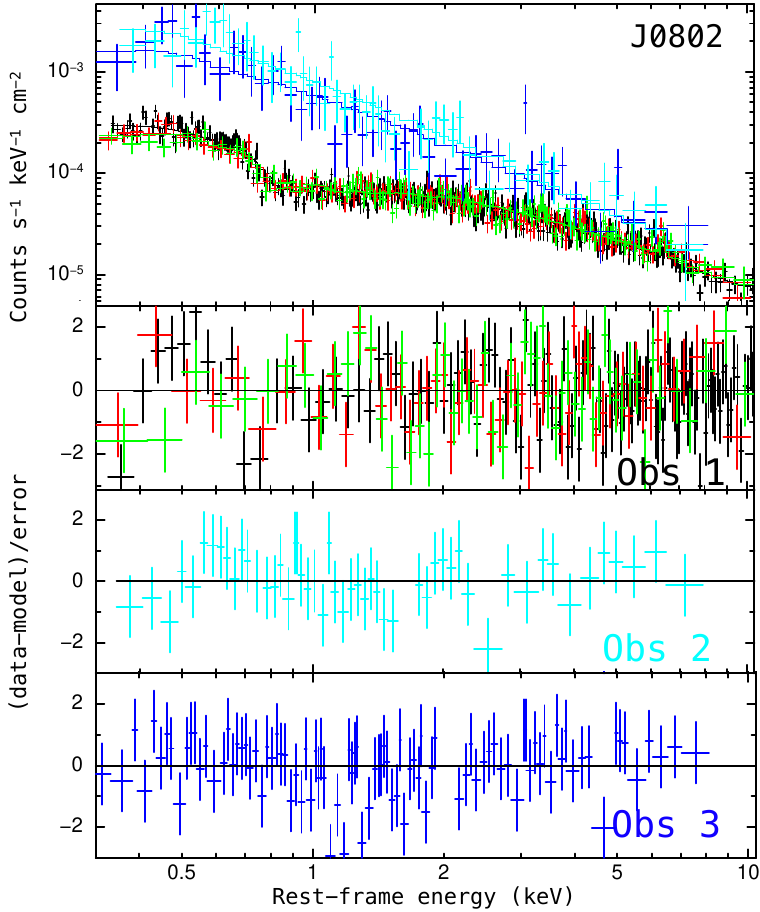}
	\caption{Top panel: \textit{XMM-Newton} EPIC (pn, black; MOS1, red and MOS2, green) and {\it Swift/XRT} (blue, cyan) spectra of J0802. The solid lines represent the best-fit models as reported in Table~\ref{tabellaJ0802}. Bottom panels: Residuals to the best-fitting model for the different observations.}
	\label{J0802fit}
\end{figure}
\label{J0802}
As shown in Figure \ref{unfoldedall}, J0802 exhibits significant spectral changes on weekly to monthly time scales. We start investigating the X-ray spectral properties of this source focusing on the two \textit{Swift/XRT} snapshots ($\sim$\,3 ks; Obs.\,2 and Obs.\,3), in which the source was found in a high-flux state. Both spectra can be adequately described with a simple power-law model absorbed by the Galactic column, as detailed in Table~\ref{tabellaJ0802} (columns 4 and 5). Such a simple phenomenological model is characterised by $C$-stat/d.o.f.\,=\,90/96 for the first {\it XRT} spectrum (Obs.\,2) and $C$-stat/d.o.f.\,=\,78/98 for the second one (Obs.\,3), taken four days later. The continuum slopes are consistent with each other, falling in a range typical for type-1 AGN, $\Gamma$\,$\sim$\,1.8--2. The source flux is statistically consistent with being constant in the hard (2--10 keV) band, but not at soft X-ray (0.5--2 keV) energies. While suggestive of a change in warm absorption (Figure \ref{unfoldedall}), at such modest data quality this can be entirely explained as a pivot effect of the continuum power law.
 A power law absorbed by the Galactic column only cannot instead provide a good representation of the \textit{XMM-Newton} spectrum taken about four months earlier (Obs.\,1), as this returns a fit statistic $C$-stat/d.o.f.~$\gtrsim$\,1.3. Prominent residuals due to ionised absorption are clearly observable in the soft X-rays, and residuals around 6.4 keV hint at the presence of a Fe-K$\alpha$ fluorescence emission line. Thus, we fit the data using the model \texttt{tbabs\,$\times$\,zxipcf\,$\times$\,(bb$\,+\,$cutoffpl\,$+$\,xillver)}. The \texttt{zxipcf} component accounts for the UTWA hosted by this AGN. The blackbody (\texttt{bb}) component is used to model the small bump in the soft X-rays, while the cut-off power law (\texttt{cutoffpl}) describes the primary continuum emission.\footnote{The narrow bandpass of {\it XMM-Newton} does not allow us to constrain the high-energy roll-over of the continuum emission. However, we use this model for consistency with \texttt{xillver}, where the illuminating continuum is described by a cut-off power law.} We assumed the latter to be the same that produces the reflected component, which is separately modelled with  \texttt{xillver} \citep[e.g.][]{Garcia2014,Dauser2016}.

In the fit, we calculated the photon index $\Gamma$ (assumed to be the same for both the primary and reflected components), as well as the ionisation parameter and column density of the warm absorber. The high-energy cut-off was kept fixed at its default value of 300 keV, as it cannot be constrained with the current data sets. For the \texttt{xillver} component, we assumed the reflecting material to be neutral ($\log \xi$\,=\,0) and to have Solar metallicity ($A_{\rm Fe}$\,=\,1), and we fitted only its normalisation. This model successfully reproduces the {\it XMM-Newton} spectrum of J0802, yielding $C$-stat\,=\,3250 for 3403 d.o.f.; the corresponding best fit and relevant quantities are shown in Figure \ref{J0802fit} and listed in Table~\ref{tabellaJ0802}. The highest-quality X-ray spectrum of J0802 is therefore consistent with a steep continuum emission ($\Gamma \sim 1.94$) modified by a UTWA with $N_{\rm H}$\,=\,(5.3$\pm$1.5)\,$\times$\,10$^{22}$ cm$^{-2}$ and $\log\xi$\,=\,1.2$ \pm$0.1 along our line of sight, implying that the dramatic variability exhibited by this source can be largely attributed to changes in the properties of the intervening material. Specifically, within just four months, the observed soft X-ray flux increased by a factor of $\sim$\,10. As a final step, we tested for the possible presence of any warm absorber also during the {\it Swift} snapshots by adding a \texttt{zxipcf} component to the best-fitting power-law. We fixed the ionisation parameter to the value found in Obs.\,1 ($\log \xi$\,=\,1.2), while allowing the column density to vary freely. This resulted in two upper limits on the column density that are more than an order of magnitude lower than the value established from the {\it XMM-Newton} spectrum, as reported in Table~\ref{tabellaJ0802}. We performed the same test but fixing $N_{\rm H}$ and computing $\log\,\xi$. This leads to a very large ionisation parameter, $\log\,\xi$\,$\sim$\,3. However the intrinsic luminosities only show little variability among the epochs, favouring a scenario where the variations are produced by $N_{\rm H}$ changes rather than an evolution in $\log\,\xi$.
\begin{table}
\centering
\caption{\small{Best-fit values obtained for J0802.}\label{tabellaJ0802}}
\resizebox{\columnwidth}{!}{%
\begin{tabular}{c c c c c}
\hline
Component & parameter & Obs.\,1 & Obs.\,2 & Obs.\,3 \\
%\hline
\texttt{tbabs} & $N_{\rm H}$ ($\times\,10^{20}$) & 4.18$\dagger$ & 4.18$\dagger$ & 4.18$\dagger$ \\
\texttt{zxipcf} & $N_{\rm H}$ ($\times\,10^{22}$) & 5.3$\pm$1.5 & $<$0.3 & $<$0.1 \\
 & $\log\xi$ & 1.2$\pm$0.1 & 1.2$\dagger$ &1.2$\dagger$ \\
 & CF & 0.96$\pm$0.02 & 1$\dagger$ & 1$\dagger$ \\
\texttt{bb} & $T_{\rm bb}$ (keV) & 0.25$\pm$0.04 & -- & -- \\
 & Norm (10$^{-5}$) & 2.0$\pm$1.5 & -- & -- \\
\texttt{cutoffpl} & $\Gamma$ & 1.91$\pm$0.25 & 1.8$\pm$0.1 & 2.0$\pm$0.1 \\
 & Norm ($10^{-4}$) & 5.4$\pm$2.0 & 5.6$\pm$0.4 & 9.4$\pm$0.6 \\
\texttt{xillver} & Norm (10$^{-6}$) & 4.6$\pm$2.4 & -- & -- \\
$K1$ & pn/mos1 & 1.02$\pm$0.03 & -- & -- \\
$K2$ & pn/mos2 & 1.04$\pm$0.03 & -- & -- \\
\hline
$F_{0.5-2\,\rm keV}$ & ($\times\,10^{-13}$) & 2.3$\pm$0.8 & 11.0$\pm$3 & 18.0$\pm$2 \\
$F_{2-10\,\rm keV}$ & ($\times\,10^{-12}$) & 1.25$\pm$0.02 & 1.9$\pm$0.4 & 2.4$\pm$0.5 \\
$\log L_{2-10\,\rm keV}$ &  & 42.82$\pm$0.01 & 42.83$\pm$0.03 & 43.03$\pm$0.02 \\
\hline
$C$-stat/d.o.f. &  & 3250/3403 & 90/96 & 78/98 \\
\hline
\end{tabular}
}% fine resizebox
\\[3pt]
\small\textbf{Notes:} The symbol $\dagger$ is used for the parameters that were kept fixed during the fitting procedure and has the same meaning in all the tables of this study. Column densities are in cm$^{-2}$; the normalisation of the blackbody component is described in \url{https://heasarc.gsfc.nasa.gov/xanadu/xspec/manual/XSmodelDiskbb.html}. The normalizations of the power-law and \texttt{xillver} components are in photons keV$^{-1}$ cm$^{-2}$ s$^{-1}$ at 1 keV. Fluxes are in erg cm$^{-2}$ s$^{-1}$, while the intrinsic (i.e. absorption-corrected) luminosities are in erg s$^{-1}$. These units apply to all the subsequent spectral-fit tables.
\end{table}

\subsection{J0809}
\begin{figure}[t]
   \centering
   \includegraphics[width=\columnwidth]{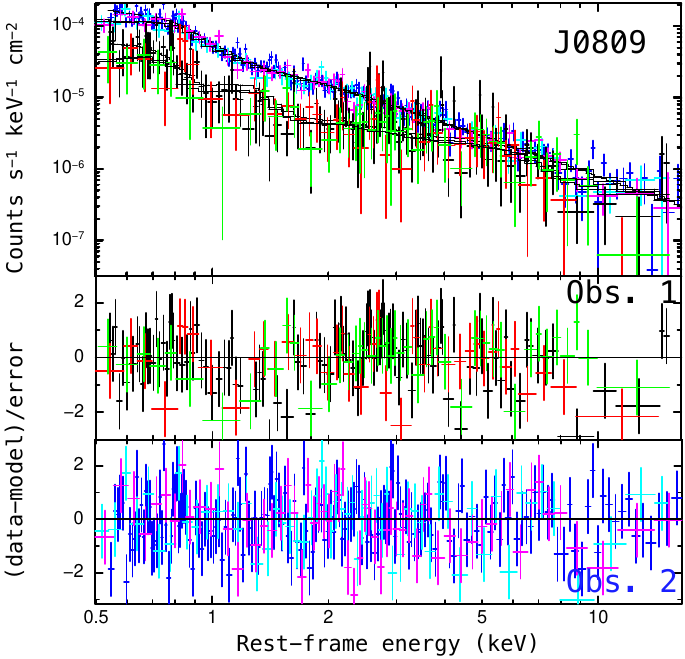}
     \caption{Top panel: \textit{XMM-Newton} EPIC spectra of J0809 (pn, black and blue; MOS1,  red and cyan; and MOS2, green and magenta). The solid lines represent the best-fit models as reported in Table~\ref{tabJ0809}.
Bottom panels: Residuals to the best-fit model for the two observations here studied. }
         \label{finaleJ0809}
\end{figure}
\begin{table}
	\centering
	\caption{\small{Best-fit values obtained for J0809.
    \label{tabJ0809}}}
	\begin{tabular}{c c c c c}
		\hline
         Component &parameter&Obs.\,1 & Obs.\,2 & \\
         \texttt{tbabs} & $N_{\rm H}$ ($\times\,10^{20}$) &3.6$\dagger$ &3.6$\dagger$\\
         \texttt{zxipcf} &$N_{\rm H}$ ($\times\,10^{22}$) &39$^{+16}_{-13}$ & $<$0.2 \\
         &$\log\xi$ & 2.80$\pm$0.08& 2.80$\dagger$ &\\
         &CF&1.0$\dagger$& 1.0$\dagger$ &\\
         \texttt{zbb}& $T_{\rm bb}$ & --&0.12$\pm$0.01 \\
         & Norm (10$^{-6}$) & --& 6.2$\pm$1.0  \\
         \texttt{zpowerlaw} & $\Gamma$ &$1.92\pm0.09$ &1.82$\pm$0.06\\
         &Norm ($10^{-5}$) &5.0$\pm$1.0 &3.7$\pm$0.2 \\
         $K1$ & pn/mos1&0.90$\pm$0.16 &0.98$\pm$0.05 \\
         $K2$ & pn/mos2&0.87$\pm$0.16 &1.02$\pm$0.04   \\
         \hline
         $F_{0.5-2\,\rm keV}$&  ($\times\,10^{-14}$) &2.6$\pm$0.4 &8.4$\pm$0.2  &\\
         $F_{2-10\,\rm keV}$&  ($\times\,10^{-14}$) &9.8$\pm$1.0 &12.5$\pm$0.7  &\\
         $\log L_{2-10\,\rm keV}$& &44.34$\pm$0.02 &44.35$\pm$0.01 &\\
         \hline
         $C$-stat/d.o.f.&&225/221 &1149/1174  \\
         \hline
%   \hline
\end{tabular}
\end{table}
\indent This broad-line AGN was first targeted by \textit{XMM-Newton} as part of a larger sample of highly accreting objects \citep[][]{Laurenti2022}. The original observation (Obs.\,1) revealed substantial suppression of photons in the rest-frame 1--2 keV energy band, as expected in the presence of ionised absorption. Approximately 2.5 years after the first pointing, J0809 was observed again with \textit{XMM-Newton}. The new observation (Obs.\,2) caught the source at a higher flux level (see Figure \ref{unfoldedall}, top-right panel), showing no obvious evidence of absorption.

To account for the evolution of the observed spectral shape, we first modelled the high-flux observation of J0809 using a simple representation consisting of a primary hard X-ray continuum (\texttt{zpowerlaw}) and a soft X-ray excess component described with a blackbody (\texttt{zbb}). Galactic absorption was also included. This model provides a fully acceptable fit, with $C$-stat/d.o.f.\,=\,1149/1174 (see Figure \ref{finaleJ0809} and Table \ref{tabJ0809}). We further tested whether the data allow for any intervening ionised absorption by adding a multiplicative \texttt{zxipcf} component with both the ionisation parameter and the column density left free to vary. This resulted in a negligible improvement to the fit ($\Delta C$-stat/$\Delta$d.o.f.\,=\,$-$5\,/$-$2), implying that no UTWA is required in this observation. Considering the possible degeneracy between the prominence of the soft emission component and the depth of the warm absorption trough, we also applied a model consisting of a simple power law modified by the \texttt{zxipcf} absorber, but this significantly worsens the fit ($\Delta C$\,=\,+65 for the same number of degrees of freedom) and suggests that a soft component is required by the data.

From the top-right panel of Figure \ref{unfoldedall}, it is clear that the blackbody plus power-law continuum model cannot reproduce the 2019 spectrum of J0809. Preliminary tests also showed that the low-flux spectrum does not require a blackbody component below 1 keV; we therefore omitted it and included instead a multiplicative \texttt{zxipcf} component. We fitted the continuum photon index and normalisation, together with the UTWA ionisation parameter and column density.
This returns a fit with $C$-stat\,=\,225 for 221 d.o.f., consistent with a primary power-law emission ($\Gamma$\,$\sim$\,1.9) absorbed by a highly ionised ($\log\xi$\,$\sim$\,2.8), thick ($N_{\rm H}$\,$\sim$\,$39 \times 10^{22}$ cm$^{-2}$) UTWA (see Table~\ref{tabJ0809}).

By adding a \texttt{zxipcf} component to the best-fitting model of Obs.\,2 (which includes the soft blackbody emission) and fixing its ionisation parameter to the value derived for Obs.\,1, we obtain an upper limit for the column density of $N_{\rm H}$\,$<$\,$0.2 \times 10^{22}$ cm$^{-2}$. We also tested the opposite scenario by freezing $N_{\rm H}$ to the best-fit value of Obs.\,1 and computing $\log\xi$. This test leads to a lower limit of $\log\xi\,>\,$3.5. Thus, the extreme change experience by J0809 in its soft X-ray flux, with a variation of about a factor of 3 in less than two years in the source reference frame, apparently followed the disappearance of the UTWA along our line of sight. This obscured to unobscured transition is similar to what has already been noticed for J0802. To the best of our knowledge, the soft X-ray ionised warm absorber detected in J0809 may represent the most distant one identified in an AGN to date. Indeed, at higher redshift it becomes difficult to determine the ionisation state of the absorbing gas, as any flux recovery at low energies progressively shifts outside the optimal bandpass of current X-ray detectors.

\begin{figure}
   \centering
   \includegraphics[width=\columnwidth]{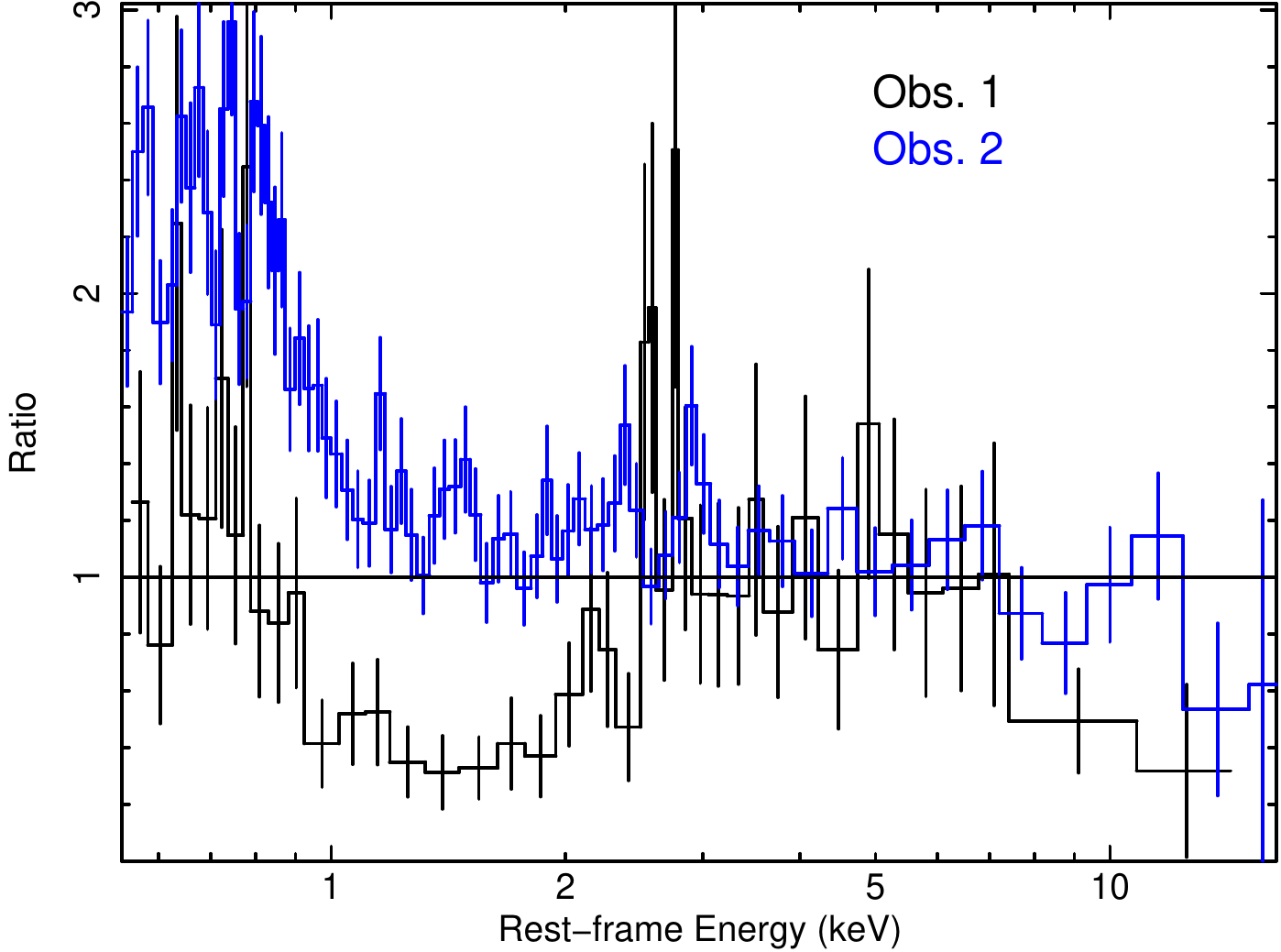}
     \caption{Ratios to a simple power law fitting the 3--5 keV band of the Obs.\,1 (black) and Obs.\,2 (blue) \textit{EPIC}-pn spectra of J0809. The spectrum of the first pointing shows significant depletion of counts in the soft and hard X-rays.}
         \label{ratiosJ0809}
\end{figure}

Interestingly, the properties of the UTWA in J0809 are usually associated with powerful UFO. As we cannot estimate the kinematics of this absorber given the limited energy resolution of the EPIC spectra and the poor S/N of the Reflection Grating Spectrometer (RGS) data, we will consider this as an extreme case of UTWA. However, by inspecting the J0809 panel in Figure \ref{unfoldedall}, one can appreciate a significant deficit of counts in the hard X-ray spectrum above $\approx$\,7 keV. This is particularly evident in the residuals of Obs.\,1, as also shown in Figure \ref{finaleJ0809}. To further examine this behaviour, we performed a joint fit of the two EPIC-pn observations in the 3--5 keV energy range using a simple power-law model. We then inspected the ratio between this model and the broadband spectra over the 0.3--10 keV range. The results of this analysis are shown in Figure \ref{ratiosJ0809}. In Obs.\,1, the data clearly reveal the presence of significant soft X-ray absorption, due to the UTWA. In addition, a prominent drop in the hard X-ray regime, possibly associated with a different absorbing component, is also observed. In contrast, Obs.\,2 shows no obvious signatures of these two absorption systems, although there are marginal indications of absorption in the hard X-ray band. This tentative deficit of counts above 7 keV might be associated to the presence of a UFO, of which the UTWA detected in the soft X-rays could represent a lower-ionisation counterpart, as already reported in other objects \citep[see, e.g.][]{Longinotti2015}. Higher quality data in the hard band are needed to confirm this conjecture about the simultaneous presence of a UTWA and a high-velocity wind.

\subsection{PG\,1535+547}
When in a high-flux state, PG\,1535+547 is the brightest source in our sample, with $F_{2-10\,\rm keV}$\,$\sim$\,$2 \times 10^{-12}$ erg cm$^{-2}$ s$^{-1}$, and it exhibits remarkable flux and spectral variability (see the corresponding panel in Figure \ref{unfoldedall}). It also has the richest data set among our sources (Section 2). In 2016, \textit{XMM-Newton} and \textit{NuSTAR} captured this AGN in an unprecedented low-flux state \citep[Obs.\,5--6;][]{MadathilPG}. As visible in Figure \ref{unfoldedall}, the 2016 spectra reveal a broad absorption trough in the soft X-ray band and a prominent Fe\,K$\alpha$ emission feature.

Given the broader energy coverage of the 2016 observations, we began characterising the spectral properties of PG\,1535+547 using these joint {\it XMM-Newton} and {\it NuSTAR} exposures. As a preliminary step, we examined a restricted portion of the spectrum (4–9 keV), which can be adequately described with a power law and a Gaussian emission line. For the narrow Fe\,K$\alpha$ feature ($\sigma$\,$<$\,110 eV), we measure an equivalent width of $\sim$\,300 eV, suggesting the presence of material with a high $N_{\rm H}$ \citep[$\sim$\,$10^{23}$ cm$^{-2}$;][]{Ghisellini1994}, compatible with the obscuring torus. We therefore applied the following broadband model:
\texttt{tbabs\,$\times$\,zxipcf}\,$\times$\,\texttt{(bb\,$+$\,cutoffpl\,$+$\,xillver)}.
The setup and meaning of each model component are the same as those described in Section \ref{J0802}. In the fit, we allowed the continuum photon index, the WA properties, and the normalisation of the reflection component to vary freely. The reflecting material was again assumed to be neutral ($\log \xi$\,=\,0) with Solar abundance ($A_{\rm Fe}$\,=\,1), while the high-energy cut-off was fixed at 300 keV.
We note that our aim here is to obtain a good phenomenological description of the observed continuum in order to properly constrain the properties of the warm absorption components. We refer to the work of \citealt{MadathilPG} for a physically motivated interpretation of the broadband emission of the source.

\begin{figure}
   \centering
   \includegraphics[width=\columnwidth]{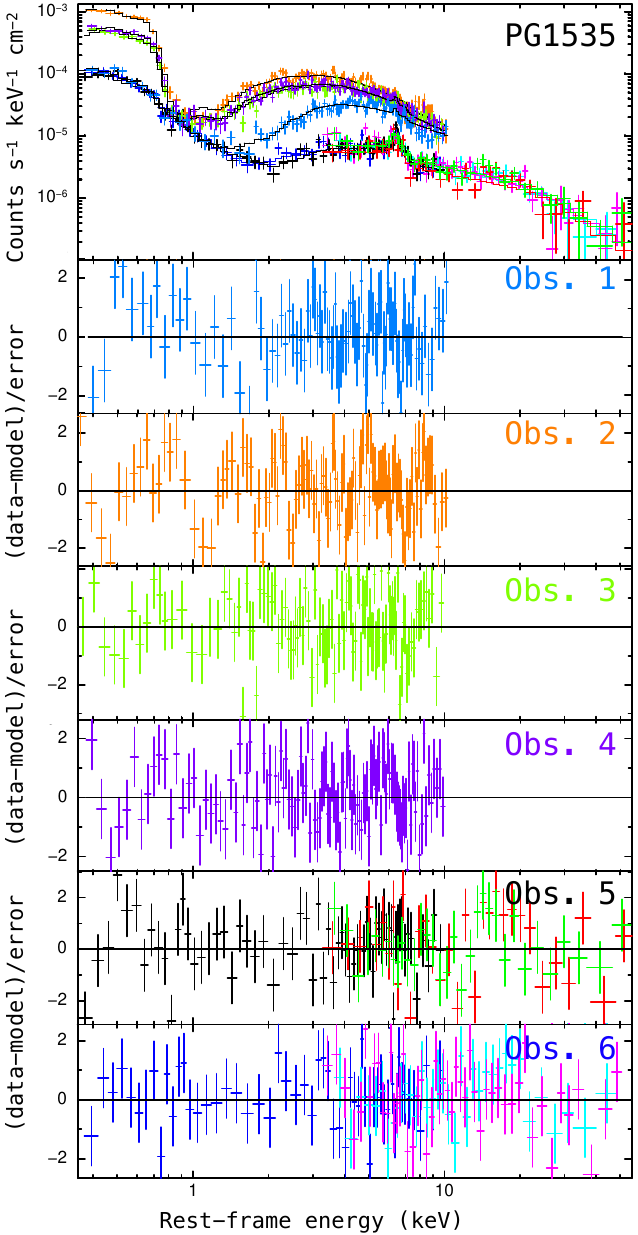}
     \caption{Top panel: \textit{XMM-Newton} \textit{EPIC}-pn spectra and best fits of PG\,1535+547. Bottom panels: Corresponding residuals.}
         \label{finalepg}
   \end{figure}

The model assumed above yields for Obs.\,5--6 the best fit reported in Table~\ref{tabellapg} and shown in Figure \ref{finalepg}, with the corresponding residuals plotted in the lower panels. These  more recent data, in contrast to previous observations reported by  \citet{Ballo2008}, do not strictly require the presence of the neutral absorber; instead, the spectrum is dominated by an ionised absorber with very high column density ($N_{\rm H}$\,$\sim$\,10$^{23}$ cm$^{-2}$) and mild ionisation ($\log \xi$\,$\sim$\,1.1), in broad agreement with earlier epochs. Remarkably, this is a key ingredient also in the more sophisticated model of \citet{MadathilPG}, even if the variability of the WAs across Obs.\,1--6 is not considered in that study.

To model the remaining spectra from Obs.\,1–4, we added a multiplicative neutral absorber to the best-fitting model described above, reaching a configuration equivalent to that used by \citet{Ballo2008}. As expected, this provides an excellent representation of all the X-ray spectra of PG\,1535+547 from 2002--2006. The best fits and corresponding parameters are shown in Figure \ref{finalepg} and listed in Table~\ref{tabellapg}. We note that fitting the continuum photon index for Obs.\,1 returns an unphysically soft value of $\Gamma$\,$>$\,3, larger than what is commonly observed in NLSy1 \citep[$\sim$\,2--2.5; e.g.][]{Brandt1997,Bianchi2009,Zhou2010,Gliozzi2020}. We therefore fixed this parameter to a value consistent with that obtained for Obs.\,5--6, since those spectra exhibit a rather similar shape. The spectra from Obs.\,2–4, taken over the course of about one week, require the presence of at least two absorbers: a neutral one with $N_{\rm H}$\,$\sim$\,$2 \times 10^{21}$ cm$^{-2}$, and a second one that is ionised ($\log \xi$\,$\sim$\,1.2) and significantly thicker ($N_{\rm H}$\,$\sim$\,$2 \times 10^{23}$ cm$^{-2}$). The underlying continuum is quite steep ($\Gamma$\,$>$\,2) in all observations, consistent with the Narrow-Line Seyfert 1 classification of PG\,1535+547. The flux associated with the reflection component appears to increase when the continuum flux is low, suggesting that the reflecting gas is located far from the SMBH and is likely cold. In all observations, a soft X-ray bump-like emission component is present with a fairly constant shape ($kT_{\rm bb}$\,$\sim$\,0.15 keV), but with varying normalisation. Given this observed variability, we tested whether a possible degeneracy exists between the blackbody normalisation and the parameters derived for the UTWA. To this end, we repeated the fits by fixing the blackbody temperature and normalisation to the values obtained from a simultaneous fit of the three 2006 observations ($kT$\,=\,0.17 keV and Norm$_{\rm bb}$\,$=$$\,1.45 \times 10^{-4}$, the latter parameter being defined as $L_{39}/D_{10}^2$, where $L_{39}$ is the source luminosity in units of $10^{39}$ erg/s and $D_{10}$ is the distance to the source in units of 10 kpc. The resulting absorber parameters are consistent, within the uncertainties, with those reported in Table~\ref{tabellapg}.

\begin{table*}
	\centering
	\caption{\small{Best-fit values obtained for all the observations of PG\,1535+547 analysed in the present work.}\label{tabellapg}}
	\begin{tabular}{c c c c c c c c}
		\hline
         Component &parameter&Obs.\,1 & Obs.\,2 &  Obs.\,3&  Obs.\,4&  Obs.\,5& Obs.\,6 \\
         \texttt{tbabs}$\dagger$ & $N_{\rm H} (\times\,10^{20}$) &1.25 &1.25 &1.25 &1.25 &1.25 &1.25\\
         \texttt{ztbabs} & $N_{\rm H} (\times\,10^{21}$) &-- &1.2$\pm$0.1 &2.0$\pm$0.05 &1.8$\pm$0.05 &-- &--\\

         \texttt{zxipcf} &$N_{\rm H}$ ($\times\,10^{22}$) &19.2$\pm$2.5 &8.4$\pm$1.3&13.8$\pm$2.9 &10.3$\pm$2.7 &12.2$^{-3.5}_{+3.1}$ &12.7$^{+2.7}_{-3.2}$\\
         &$\log\xi$ &1.3$^{+0.3}_{-0.1}$ &1.2$\pm$0.1&1.45$\pm$0.05 &1.2$\pm$0.2 &1.1$\pm 0.1$ &1.20$\pm$0.05\\
         &CF &$>$\,0.99 &$>$\,0.99&$>$\,0.99 & $>$\,0.99 &$>$\,0.99&$>$\,0.96\\

         \texttt{bb}& $T_{\rm bb}$ &0.90$\pm$0.02 &0.18$\pm$0.01 &0.15$\pm$0.03 &0.17$\pm$0.01 &0.15$\pm$0.02 &0.18$\pm$0.04\\
         & Norm (10$^{-5}$)& 2.0$\pm$0.5&30$\pm$25&20$\pm$5 &62$\pm$40 &1.4$\pm$0.7 &2.2$\pm$0.7\\

         \texttt{cutoffpl}& $\Gamma$ &2.06$\dagger$&2.34$\pm$0.22 &2.5$\pm$0.4 &2.50$\pm$0.35 &2.03$\pm$0.17 &2.08$\pm$0.16\\
         & Norm ($10^{-4}$) &12$^{+2}_{-4}$ &36$\pm$7 &40$\pm$7&39$\pm$15 & 1.6$\pm$0.3&2.3$\pm$0.8 \\
         \texttt{xillver}& Norm ($10^{-6}$) &$<$\,1.35&$<$\,4.2 &$<$\,1.7 &$<$\,3.6 &13$\pm$5 &11$\pm$5\\
         $K$ & FPMA& --& --& -- & -- &0.93$\pm$0.07&1.18$\pm$0.11\\
         $K$ & FPMB& --& --& --  &-- &1.17$\pm$0.06&1.2$\pm$0.11\\
   \hline
         $F_{0.5-2\,\rm keV}$&   ($\times\,10^{-14}$) &4.2$\pm$0.9 &30$\pm$5 &16$\pm$1 &19$\pm$6 & 3.3$\pm$0.4&3.3$\pm$0.3 \\
         $F_{2-10\,\rm keV}$&   ($\times\,10^{-13}$)  &15$\pm$1 &30$\pm$2 &23$\pm$3 &24$\pm$5 &3.7$\pm$0.2&3.6$\pm$0.3\\
         $\log L_{2-10\,\rm keV}$ & &43.14$\pm$0.01 &43.30$\pm$0.01 &43.27$\pm$0.01 &43.26$\pm$0.02 &42.36$\pm$0.02& 42.37$\pm$ 0.02 \\
          \hline
          $\chi^2$/d.o.f. & &127/131 &142/107 &118/97 &125/110  &164/130 & 133/132\\
          \hline
 %         \hline
\end{tabular}
\end{table*}
It is worth noting that our results on the complex absorption structure of this source are in qualitative agreement with the findings of \citet{MadathilPG}. Despite their more refined spectral modelling and the adoption of custom XSTAR \citep{Kallman2001} grids to model the multiple absorbers in this source, the authors also report on ionised absorbers with high column densities ($N_{\rm H}$\,$\sim$\,$1\times10^{23}$ cm$^{-2}$). This agreement between our phenomenological modelling and their physically-motivated treatment reinforces the presence of UTWA components in the circumnuclear environment of PG\,1535+547, which are relatively independent of the specific continuum and absorption modeling adopted.

\begin{figure}
\centering
\includegraphics[width=\columnwidth]{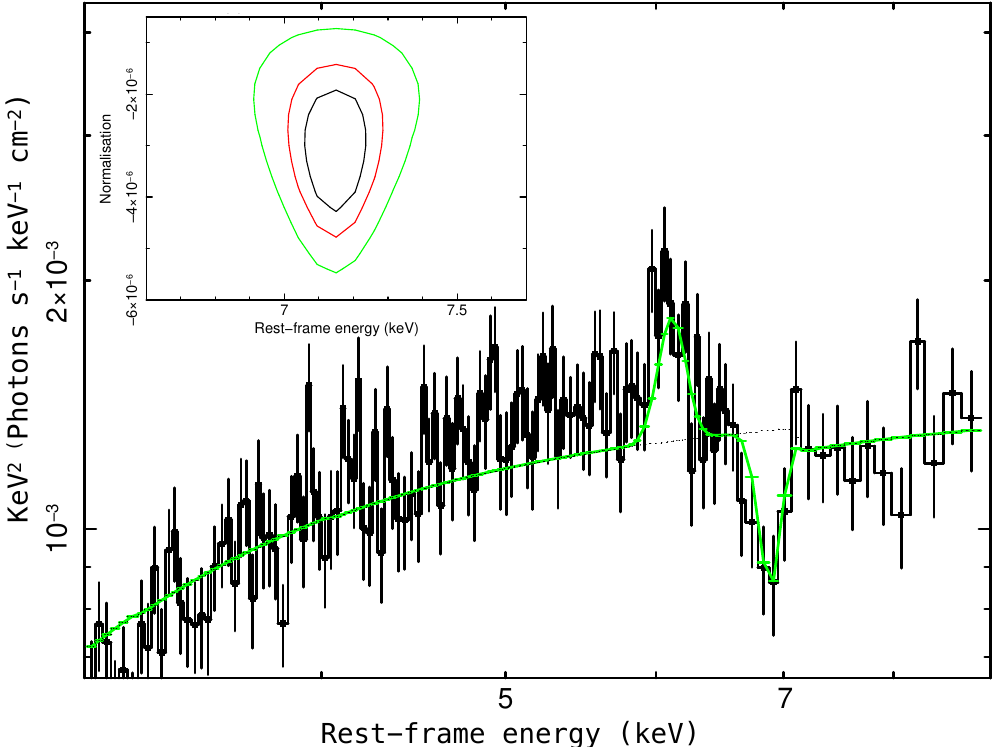}
\caption{Zoom in the 3--9 keV merged (Obs.\,2, 3 and 4) spectrum. Inset: Absorption feature detected in these combined 2006 observations of PG\,1535+547. The confidence contours refer to the 68\%, 90\% and 99\% significance levels.}
\label{discwind}
\end{figure}

Similar to the case of J0809, also in PG\,1535+547 we observe some hints of absorption features in the hard X-ray band (see Figure \ref{unfoldedall} and residuals in Figure \ref{finalepg}), especially in Obs.\,2--4 (i.e. in the high state). We focused on the three 2006 observations to constrain the properties of any possible ionised absorber responsible for the spectral drop above 6~keV.
These exposures show moderate flux variability below $\sim$\,2~keV, while their hard X-ray emission remains relatively stable. We then merged the spectra from Obs.\,2--4 using the standard \textsc{sas} task \texttt{epicspeccombine}, and analysed the resulting 2--10~keV emission. The best-fit model derived for the single spectra, in which no Fe-K absorption is included, leaves strong residuals around $\sim$\,7.1~keV (rest-frame) when applied to the merged spectrum. As shown in Figure \ref{discwind}, these residuals are highly significant and can be effectively modeled by a single Gaussian absorption line with a fixed width of $\sigma$\,=\,100~eV.  The inclusion of this component improves the fit by $\Delta\chi^2$\,=\,$-$20 for two additional degrees of freedom. The best representation of this Gaussian line is obtained for an energy centroid $E$\,=\,7.15$\pm$0.10 keV, a normalisation of ($-$3.1$\pm$1.2)\,$\times$\,10$^{-6}$ photons cm$^{-2}$ s$^{-1}$, and an equivalent width EW\,=\,$-$100$\pm$40 eV. Interestingly, if we assume this line to be either blueshifted Fe\,\textsc{xxv} or Fe\,\textsc{xxvi}, we derive $v_{\rm out}$\,$\sim$\,30,000 or $v_{\rm out}$\,$\sim$\,19,000 km s$^{-1}$, respectively, corresponding to a significant fraction of the speed of light (10\% and 6\%), thus placing this absorber in the UFO regime. The clear detection of an absorption feature in the combined spectrum suggests that the associated disc wind is persistent on timescales of at least one week, as any feature exhibiting more rapid variations in depth and/or energy would be smeared out in the co-added data.

\begin{table*}
	\centering
	\caption{\small{Best-fit values obtained for all the observations of J0232 analysed in the present work.}\label{tabellaleda}}
	\begin{tabular}{c c c c c c}
		\hline
         Component &parameter&Obs.\,1 & Obs.\,2 &  Obs.\,3 & Obs.\,4\\
         \texttt{tbabs}$\dagger$ & $N_{\rm H} (\times\,10^{20}$) &7.8 &7.8 &7.8 &7.8\\
         \texttt{tbabs} & $N_{\rm H} (\times\,10^{21}$) &-- &-- &8.5$\pm$3.0 & $<$\,8.3\\
         \texttt{zxipcf} &$N_{\rm H}$ ($\times\,10^{22}$) &6.6$^{+0.5}_{-1.0}$ & 6.5$\pm$0.5&2.95$^{+0.90}_{-0.05}$& 7.8$^{+0.4}_{-1.0}$\\
         &$\log\xi$ & $-$0.55$^{+0.20}_{-0.45}$&$-$0.5$\pm$0.2&0.80$^{+0.30}_{-0.15}$&1.10$^{+0.15}_{-0.55}$\\
         &CF&0.96$\pm$0.01 & 0.98$\pm$0.01 & $>$\,0.98&$>$\,0.98\\
         \texttt{bb}& $T_{\rm bb}$ &--&--&0.17$\pm$0.01&0.20$\pm$0.05 \\
         & Norm (10$^{-5}$)&--&--&9.2$\pm$6.0&37$\pm$5\\
         \texttt{cutoffpl}& $\Gamma$ & 1.9$\dagger$&1.9$\dagger$ &1.9$\pm$0.1&1.94$\pm$0.10\\
         &Norm ($10^{-4}$) &3.6$\pm$0.7 &4.1$\pm$0.7 &9.0$\pm$0.2 &9.1$\pm$1.4\\
         \texttt{xillver}&Norm (10$^{-6}$)&$<$\,9.5&8.4$\pm$4.8&4.6$\pm$2.4 &7.6$\pm$3.6\\
         $K1$ & pn/mos2&0.91$\pm$0.12&1.16$\pm$0.12&1.04$\pm$0.03 &--  \\
         $K2$ & pn/FPMA&--&--&1.38$\pm$0.08&-- \\
         $K3$ & pn/FPMB&--&--&0.77$\pm$0.06&-- \\
         \hline
         $F_{0.5-2\,\rm keV}$&  ($\times\,10^{-13}$) &0.27$\pm$0.07&0.23$\pm$0.04&3.2$\pm$0.4&2.1$\pm$0.2\\
         $F_{2-10\,\rm keV}$&  ($\times\,10^{-13}$) &6.1$\pm$0.7&7.3$\pm$0.5&23.1$\pm$0.2&23.6$\pm$0.1\\
         $\log L_{2-10\,\rm keV}$& &42.35$\pm$0.04& 42.50$\pm$0.02&42.88$\pm$0.01&42.90$\pm$0.01\\
         \hline
         $\chi^2$/d.o.f.&&243/240&240/260 &623/579& 330/375\\
         \hline
   \hline
\end{tabular}
\end{table*}

\subsection{J0232}
\begin{figure}
   \centering
   \includegraphics[width=\columnwidth]{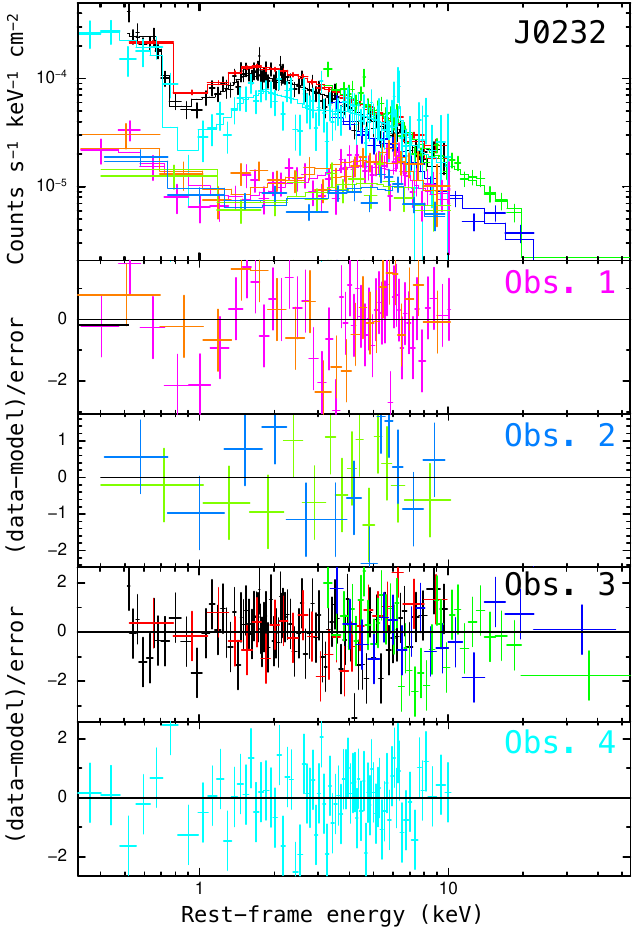}
     \caption{Top panel: spectra and best fits of J0232, as described in Section 3.4. Bottom panels: residuals to the best-fit models. Multiple colours are used to distinguish among the different spectra obtained either with the EPIC camera or the two detectors on the focal plane of \textit{NuSTAR}.}
         \label{finaleleda}
\end{figure}
We began by modelling the joint \textit{XMM-Newton} and \textit{NuSTAR} observations taken in 2021. This specific dataset was selected because the source was caught in a high-flux state, which, combined with the broader {\it NuSTAR} energy coverage, allows a more robust and detailed description of the intrinsic continuum shape.

Within \textsc{XSPEC}, we fitted the 0.3--45 keV spectra of J0232 using the model \texttt{tbabs\,$\times$\,zxipcf}\,$\times$\,\texttt{(bb\,$+$\,cutoffpl\,$+$\,xillver)}. In the fit, we determined the continuum slope, linking its value across all spectra and between the power-law and reflection components. The reflection component was treated as in the previous cases: we assumed it to originate from neutral material ($\log \xi$\,=\,0) with Solar abundance ($A_{\rm Fe}$\,=\,1). Both $\log\xi$ and $N_{\rm H}$ were left free to vary for the UTWA. This model yields a fit with $\chi^2$/d.o.f.\,=\,645/580. Some residuals remain below 1~keV, possibly indicating the presence of additional neutral absorbing material. We therefore added an extra neutral absorber (\texttt{ztbabs}), which significantly improved the fit (at the $\gg$\,3$\sigma$ level), resulting in $\chi^2$/d.o.f.\,=\,623/579. Figure \ref{finaleleda} shows the corresponding best-fit model for Obs.\,3, and Table~\ref{tabellaleda} lists the derived physical parameters. We then applied the same model to Obs.\,4, taken in 2023, during which the source is still observed at a high flux level. The model reproduces the spectral shape well. Interestingly, in this observation the  inclusion of an additional neutral absorption layer returns an upper limit on $N_{\rm H}\,<\,8.3\times10^{21}$ cm$^{-2}$ (see Table~\ref{tabellaleda}. We next applied the model to the earliest \textit{XMM-Newton} observations (Obs.\,1--2 in Table~\ref{log}, taken over consecutive orbits). In these data, no neutral absorption in excess to the Galactic value is required. Moreover, no additional soft component is needed (thus the blackbody was removed from the model), and both observations return very hard photon indices ($\Gamma$\,$\sim$\,1.3). We therefore assumed the continuum to have the same photon index as in the 2021 observation (which, thanks to {\it NuSTAR}, benefits from a broader bandpass) and re-fitted the spectra. This yields the best fits shown in Figure \ref{finaleleda}, with the corresponding parameters listed in Table~\ref{tabellaleda}. Remarkably, the continuum emission in these early observations is attenuated by an absorber with a column density comparable to that found more than a decade later for the UTWA ($N_{\rm H} \sim\,6$–$8 \times 10^{22}$ cm$^{-2}$), but with a significantly lower ionisation state ($\log \xi$\,$\sim$\,$-0.5$, compared to $\sim$\,1.1).

\section{Discussion}
In the following, we discuss our findings in the broader context of X-ray ionised absorbers in AGN, and look into the possible origin of UTWAs also considering previous studies from the literature.

\subsection{UTWA region in the $\log N_{\rm H}$--$\log \xi$ plane}

The X-ray spectroscopic analysis of the four active galactic nuclei presented here revealed the presence, in at least one epoch, (see Figure \ref{contoursall} in Appendix B), of a thick ($\log N_{\rm H}$\,$>$\,22.5) and moderately ionised ($\log \xi$ in the range 0.5--2.5) absorber, which we define as a UTWA. Our study is based on a total of 18 archival observations including {\it XMM-Newton}, {\it Swift}, and {\it NuSTAR} data, which suggest that these absorbers significantly vary in column density (and, possibly, ionisation state) on time scales as short as a few months. Here we complement the results presented in Section 3 with a discussion of other AGN in which the presence of a UTWA has been reported in published works. Specifically, these are 1E\,0754.6+3928, 3C\,445, ESO\,323$-$77, IRAS\,09149$-$6206, NGC\,985, Mrk\,704, PG\,1114+445, and Mrk\,304. We provide more details about the latter sources and their UTWAs in Appendix~\ref{Appendix}.

\begin{figure}
   \centering
   \includegraphics[width=\columnwidth]{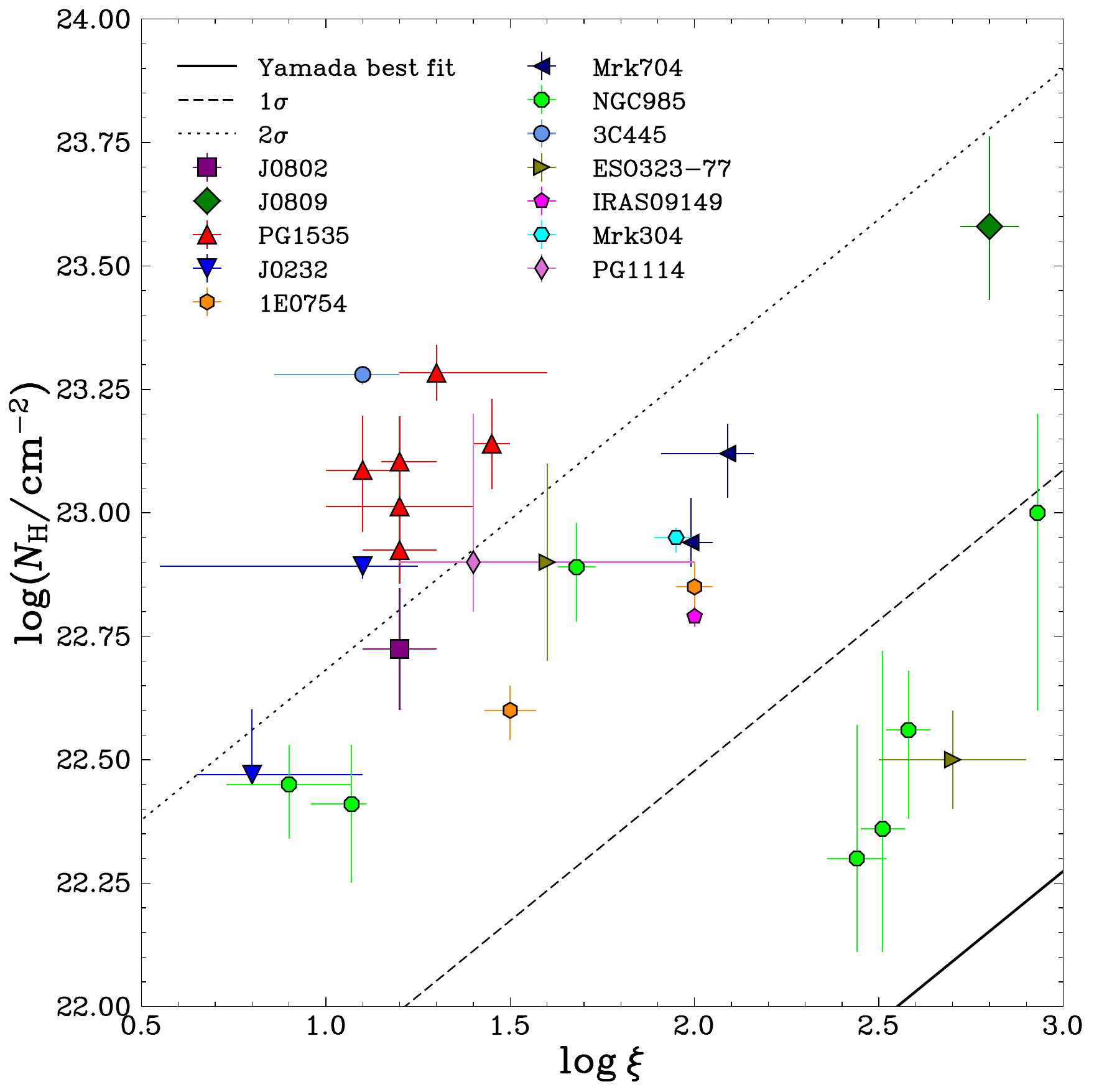}
     \caption{Properties of the UTWAs in the sources analysed in this work and the AGN from the literature listed in Table~\ref{AppT}. For the sake of comparison, we also display the global best-fit relation derived from all the X-ray absorbers in the sample of \citet{Yamada2024} as a solid black line (bottom-rigth corner), together with the 1$\sigma$ and 2$\sigma$ confidence bands.}
         \label{Yamada2024comp}
\end{figure}

Figure \ref{Yamada2024comp} shows the properties of all the UTWAs discussed or considered in this work, and compare their location in the $\log N_{\rm H}$--$\log \xi$ plane with the bulk of the sample assembled by \citet{Yamada2024}.
By definition, all the identified ultra-thick and moderately ionised absorbers populate a poorly sampled region of the parameter space of interest. While the present analysis enlarges by $\approx$\,50\% the number of objects known to host a UTWA, and the exact boundaries in both $\xi$ and $N_{\rm H}$ of this class of absorbers are somewhat indistinct (and, to some extent, model-dependent), UTWAs remain definitely rare with no obvious observational reason. This can have different physical explanations. The simplest yet still plausible one is purely geometrical, involving a somehow peculiar line of sight through which we are observing these sources. In this framework, ultra-thick matter close enough to the ionising continuum would be present in most AGN, but we can observe it only along a specific line of sight. However, as the objects harbouring UTWAs are type 1--1.5 AGN, the viewing angle to their central regions is likely close to polar, or moderately inclined at most, making the geometrical explanation hard to envisage.
Without any reliable constraint on their kinematics (at least when, as usually occurs, no high-resolution data of sufficient quality are available), we cannot even conclude that UTWAs are a genuine manifestation of the standard family of WAs, simply displaying more extreme properties possibly because forming further in than the majority of the population, i.e. somewhere in between the broad-line region (BLR) and the inner part of the torus. At this stage, then, more informative clues on the nature of UTWAs can only derive from their variability behaviour and their relation to other types of absorbers.

\subsection{Variability}
The soft X-ray band of all the sources analysed here presents strong spectral changes that are primarily driven by absorption variability (see Figure \ref{unfoldedall}). Similar strong variability is also observed in the X-ray spectra of most of the sources mentioned in Appendix A (see below and their corresponding references).
A possible explanation for these variations, which result in soft X-ray flux changes by up to a factor of $\sim$\,10, would involve intervening BLR clouds, which are typically associated with X-ray eclipses (i.e. on/off events) on short timescales. These clouds, which are generally assumed to be neutral as they completely obliterate the nuclear soft X-ray emission \citep[e.g.][]{Risaliti2007}, %(generally neutral or lowly ionised ($\log \xi\,<\,1$),
may obscure the X-ray source while crossing our line of sight along their orbits at different distances from the SMBH.

One of the most comprehensive searches for discrete X-ray absorption events in nearby AGN was performed by \citet{Markowitz2014} based on RXTE data. The authors found that eclipses in type-1 AGN typically last less than $\sim$\,100 days and are associated to modestly ionised clumps at a distance consistent with the outer portions of the BLR (or the inner radius of the dusty torus). One notable example is the detection of a $\sim$\,100-day transient absorption event in the nearby Seyfert-1 galaxy NGC\,3227, associated to a cloud with $\log \xi$\,$\sim$\,0 and $N_{\rm H}$\,=\,$6\times10^{22}$ cm$^{-2}$ located in the BLR, reported by \citet{Lamer2003}. A transition from obscured to unobscured states, or vice versa, on time scales
compatible with the transit of BLR clouds has been observed in the spectra of J0802 and J0809, although a much finer sampling of the temporal and spectral evolution is required to confirm a similar scenario. Moreover, the physical properties of the UTWA detected in J0809 appear to be rather extreme, due to the large value of the ionisation parameter of the absorbing gas. Alternatively, a persistent absorber might be present, which simply changes its opacity in response to changes in the intensity of the illuminating continuum. Indeed, in both J0802 and J0809 the UTWA appears in the low-flux state, suggesting that the source is not able to ionise the foreground gas to make it sufficiently transparent. Even neglecting any time delay, however, the variability of the intrinsic continuum, as revealed by the intrinsic 2--10 keV luminosities, is not so large to fully justify the observed effects.
On the other hand, in PG\,1535+547 and J0232 warm absorption is systematically observed over a time span of more than a decade, which seems to be at odds with an origin in terms of BLR-cloud crossing.
In the first two observations of J0232, the soft X-rays of this AGN are absorbed by matter characterised by a thick column density but a very low ionisation parameter. No significant changes are observed between these two exposures, which were taken two days apart. Conversely, in the subsequent exposures (taken after approximately 12 and 14 years), the spectral shape of the source in the soft band underwent noticeable changes (see Figure \ref{unfoldedall}, bottom right panel). The absorption troughs appear less smooth and with a sharper edge around $\sim$\,0.7 keV, with differences between Obs.\,3 and 4. Notably, the same behaviour is seen in the higher S/N spectra of PG\,1535+547. In the case of J0232, this translates into the presence of a UTWA in both Obs.\,3 and 4, which is found to vary in column density and ionisation state.
It is also worth mentioning that in these last two exposures, we detected the presence of additional absorption from neutral matter, which can possibly suggest the presence of multiple layers of gas constituting a clumpy wind.
Similarly to J0232, also PG\,1535+547 exhibits persistent signatures of thick ($\log N_{\rm H}$\,$\gtrsim$\,23) ionised absorption. Since its first observation (Obs.\,1), this source has shown a conspicuous deficit of counts with a minimum between $\sim$\,1--2 keV. In subsequent observations (Obs.\,2--4), the absorbing matter was found to be characterised by a larger column density and ionisation parameter. These four exposures were originally studied in depth by \citet{Ballo2008}, who attributed most of the variability to the warm absorbers, which change their physical properties on daily to yearly timescales. Moreover, the authors noted that strong variability largely limited the X-ray band, in contrast to the more constant optical emission of PG\,1535+547. Later, in Obs.\,5--6, this AGN was caught in an unprecedented low-flux state characterised by a prominent Fe K$\alpha$ emission line. This emission line likely originates from cold matter with a $N_{\rm H}$ exceeding $10^{23}$ cm$^{-2}$, and becomes clearly visible when the source is in a low-flux state. We note a drop of about a factor of 10 between Obs.\,2 and Obs.\,5 in the 2-10 keV luminosity. On the other hand, the UV luminosities derived from the UV filters of the Optical Monitor onboard {\it XMM-Newton} do not show this remarkable drop, suggesting that PG\,1535+547 might be entering an X-ray weak phase. Yet, this does not seem to affect the properties of the UTWA, which remain relatively stable across all epochs. This suggests that the absorbing gas could be co-spatial with the cold reflecting structures, e.g. the obscuring torus.

Summarising, with currently available observations, we cannot assess whether UTWAs have a preferred location as they can be laying from sub-pc scales and up to several pc of distance from the central black hole. This is also confirmed by literature results. In some cases (NGC\,985, \citealt{Ebrero2021}; Mrk\,704, \citealt{Matt2011}; IRAS\,09149$-$6206, \citealt{Walton2020}), the UTWA is implied to partially cover the X-ray source, suggesting a clumpy and/or filamentary structure and a relative vicinity to the primary source (see also the discussion in \citealt{Braito2011} for 3C\,445). Even so, while the geometrical covering factor can significantly evolve over time, variations in the physical parameters ($N_{\rm H}$ and $\xi$) tend to be mild over yearly timescales. Interestingly, also thanks to high-resolution data, in NGC\,985 the partially covering UTWA component is interpreted as an obscuring wind, which produces a long-lasting eclipsing event that shields the other absorbing layers located further out.

\subsection{The possible co-existence with other wind components}

As a more sophisticated and intriguing interpretation, UTWAs could represent a component of a complex, possibly multi-phase obscuring system. The recent \textit{XRISM/Resolve} observation of the prototypical UFO in the quasar PDS\,456 \citep{Xrism2025} has revealed that the disc wind in this source is constituted by myriads of clumps, indicating that the outflowing material is far from homogeneous. In addition, high-velocity soft X-ray absorbers have been detected in some AGN \citep[e.g.][]{Mehdipour2017,Kriss2018}, showing lower column densities and ionisation parameters compared to typical UFOs, yet with comparable $v_{\rm out}$, i.e. about two orders of magnitude higher than that of classical WAs (usually a few hundreds km s$^{-1}$).

In this framework, the detection of a clear UFO feature in the stacked spectrum of PG\,1535+547 from Obs.\,2--4 (Figure \ref{discwind}) is definitely remarkable. Moreover, also J0809 displays a tentative deficit with respect to the X-ray continuum emission in the Fe-K energy range (see Fig.\, \ref{ratiosJ0809}), although the presence of a disc wind cannot be safely established with the current data. Considering also the results from the literature, the possible connection between UTWAs and other wind phases, whether ultra-fast or not, becomes even more compelling. \citet{Middei2020} discussed the possible presence of a high-velocity component in the spectra of 1E\,0746+3928, while a high $\xi$ absorber with outflow velocity approaching 10,000 km s$^{-1}$ was found by \citet{Walton2020} in IRAS\,09149-6206. Some UTWAs might therefore represent the soft X-ray counterpart of disc winds or be physically associated to the UFO phenomenon itself, as suggested for the case of PG\,1114+445 \citep[e.g.][]{Serafinelli2019}, where a fast outflow with moderate ionisation appears to arise from ambient clouds entrained by the faster, more highly ionised UFO. UTWAs could trace slower or denser portions of a more extended outflow, composed of material originally ejected from the accretion disc and subsequently interacting with the surrounding medium. This scenario is qualitatively supported by the detection of absorption features in the Fe-K band of virtually all the AGN showing UTWAs (see \citealt{Jimenez2008} for ESO\,323$-$77), which may indicate the co-existence of both highly ionised and moderately ionised gas phases within the same outflow structure, whereby the former provide the pressure confinement to the latter \citep{Sanfrutos2016}.

\section{Summary and conclusions}

\indent In this work, we reported on the presence and  the properties of the ultra-thick warm absorbers observed in a sample of 12 nearby AGN. These absorbers are characterised by an ionisation degree compatible with that of standard warm absorbers (0.5\,$<$\,$\log \xi$\,$<$\,2.5), but accompanied by uncommonly high column density ($\log N_{\rm H}$\,$>$\,22.5), from which follows our definition (see Figure \ref{Yamada2024comp}). For four of these sources, we performed a dedicated spectroscopic analysis, while for the remaining ones we adopted parameters reported in the literature. All these sources exhibit extreme variability in the soft X-ray band, which can be ascribed to absorption variability.

\indent The physical origin of UTWAs and their relation with other X--ray detected ionised absorbers remain unclear, especially considering their apparent rarity compared to the results of recent ensemble studies \citep[][]{Yamada2024}. It is possible that UTWAs correspond to transient or geometrically constrained phases of the absorbing material, observable only under specific lines of sight \citep[e.g. Mrk\,704,][]{Matt2011}. Alternatively, they may represent a denser or less ionised phase of a more complex, multi-phase outflow, potentially associated with accretion disc winds (e.g. PG\,1535+547 and J0809; see Sect. 4.3). In any case, given their extreme values of $N_{\rm H}$ and $\xi$, UTWAs deserve detailed investigation, as their inclusion is essential for a comprehensive characterisation of ionised absorbers and nuclear outflows in AGN.

\indent The general lack of high-quality, high spectral resolution data, coupled with the uneven temporal sampling of our data sets, leaves several questions concerning these absorbers unanswered. Specifically, the lack of accurate information on their kinematics poses a major obstacle to fully understanding them. By providing high-resolution spectra in the critical range $\sim$\,1--3 keV, {\it XRISM} would have represented a major step forward in both the demography and physical understanding of UTWAs; however, the gate valve issue has so far prevented any substantial progress with this respect. It is likely that a truly in-depth, high-resolution characterisation of these elusive, thick, ionised absorbers, capable of unveiling their detailed physical properties, will therefore have to await a next-generation mission like \textit{New Athena} \citep[][]{Cruise2025}. On the other hand, well-designed observing campaigns using existing X-ray observatories will be crucial for better characterising the variability properties of UTWAs and their relation to different X-ray flux states, while deep observations will help clarify the possible connection between UTWAs and UFOs.

\begin{acknowledgements}
We are grateful to the referee for their valuable comments and suggestions. RM acknowledges financial support from the INAF Scientific Directorate. AT acknowledges financial support from the Bando Ricerca Fondamentale INAF 2022 Large Grant `Toward a holistic view of the Titans: multi-band observations of $z>6$ QSOs powered by greedy supermassive black holes' and from the Bando Ricerca Fondamentale INAF 2024 Large Grant `The DEepest study of LUminous QSOs in X-ray at $z$\,=\,2--7'. EP acknowledges financial support from PRIN-MUR-2022 grant “Advanced X-ray modeling of black hole winds” (DRAGON; No. PRIN 2022K9N5B4) and Bando Ricerca Fondamentale INAF 2023, L.P. 1.05.23.01.06 (“The XRISM-to-XIFU (X2X) Agreement and Beyond"). ATF was supported by an appointment to the NASA Postdoctoral Program at the NASA Goddard Space Flight Center, administered by Oak Ridge Associated Universities under contract with NASA. This work relies on archival data, software or online services provided by the Space Science Data Center\,--\,ASI, and it is based on observations obtained with XMM-Newton, an ESA science mission with instruments and contributions directly funded by ESA Member States and NASA.

\end{acknowledgements}

\bibliographystyle{aa}
\bibliography{ultrathickv1}
\clearpage
\begin{appendix}
\section{AGN with known UTWAs from literature}
\label{Appendix}

In addition to the four AGN presented in Section 2.1, we consider in this paper eight other objects that have been revealed to exhibit an X-ray absorber whose column density and  ionisation parameter are consistent with our definition of UTWA in at least one X-ray observation. Five of them are already included in \citet{Yamada2024} (i.e. 1E\,0754.6$+$3928, NGC\,985, 3C\,445, ESO\,323$-$77, and IRAS\,09149$-$6206),  and we summarise their properties and classification in Table A.1, while the characteristics of their UTWAs are listed in Table A.2 as reported in the discovery papers. We note that two additional sources occupy the UTWA region in the compilation of \citet{Yamada2024}, but are not considered here: PDS\,456 and NGC\,4051. In PDS\,456, the absorber with high column density and moderate ionisation can be clearly identified as a components of the multi-phase ultra-fast outflow \citep{Reeves2020}, while in NGC\,4051 the column density strongly depends on whether the absorber is photo- or collisionally ionised \citep{Ogorzalek2022}.

The four remaining objects in our extended sample are Mrk\,704, Mrk\,304, PG\,1114+445. We provide the basic details of all the sources below, and list the properties of the UTWAs detected in their X-ray spectra in Table A.1.

\noindent \textbf{1E\,0754.6+392.8} is a local ($z$\,=\,0.096) radio-quiet Narrow Line Seyfert~1 galaxy with a BH mass of $\log(M_{\rm BH}/M_\odot)$\,=\,8.15, and a bolometric luminosity of $\log(L_{\rm bol}/{\rm erg\,s^{-1}})$\,=\,45.4 \citep[][]{Berton2015,Sergeev2007}. It is the brightest among the AGN in the {\it NuSTAR} serendipitous source catalogue \citep{Lansbury2017}. The analysis of the X-ray data from \textit{XMM-Newton}, \textit{NuSTAR}, and \textit{Swift/XRT} by  \citealt{Middei2020} revealed a persistent UTWA with $N_{\rm H}$\,$\sim$\,4--8\,$\times 10^{22}$ cm$^{-2}$, and mildly variable ionisation (1.5\,$\lesssim$\,$\log \xi$\,$\lesssim$\,2.0), compatible with a UTWA. Moreover, the authors also report on the possible presence of an additional high-ionisation, high-velocity absorber with UFO-like properties.

\noindent \textbf{NGC\,985} is a nearby luminous \citep[$L_{\rm 2-10\,keV}$\,$\sim$\,$4.2 \times 10^{43}$ erg s$^{-1}$][]{Bianchi2009} Seyfert~1 galaxy located at $z$\,=\,0.043, hosting a supermassive BH with $\log(M_{\rm BH}/M_\odot)$\,$\sim$\,8.05 \citep[][]{ONeill2005}. \citealt[][]{Ebrero2021} analysed its rich {\it XMM-Newton}/RGS spectra, revealing the presence of up to four warm-absorption components, including UTWAs, with $N_{\rm H}$ and $\log\xi\,\sim\,10^{21}$--10$^{23}$\,cm$^{-2}$ and $\sim$\,1--3, respectively, and outflow velocities of $\sim$\,350--5,100 km s$^{-1}$. A further, mildly ionised and partially covering component is identified with an obscuring wind that appears to shield the other absorbers from the ionising continuum \citep[][]{Ebrero2021}.

\noindent \textbf{3C\,445} is a local ($z$\,=\,0.057) Broad Line Radio Galaxy (BLRG) \citep[][]{Kronberg1986}. \citet{Bettoni2003} reports the mass of the central SMBH to be $M_{\rm BH}$\,$\sim$\,$2 \times 10^8~M_\odot$ while a luminosity $L_{\rm Bol}$\,$\sim$\,$2\times 10^{45}$ erg s$^{-1}$ was inferred by \citet[][]{Marchesini2004}. The primary X-ray continuum is strongly obscured by a UTWA with $N_{\rm H}$\,$\sim$\,2--3\,$\times10^{23}$ cm$^{-2}$ and $\log\xi$\,$\sim$\,1, possibly varying over yearly time scales, and interpreted as an equatorial disc wind located within the molecular torus \citep[][]{Braito2011}.
%\item

\noindent \textbf{ESO\,323--G77} lies at $z$\,=\,0.015 and is known to be a polar-scattered Seyfert~1.2 galaxy \citep[][]{Veron-Cetty2006}. \citet{Wang2007} reported on this source being powered by a SMBH with $M_{\rm BH}$\,$\sim$\,$2.5\times10^{7}~M_\odot$, while \citet{Serafinelli2023} derives a source bolometric luminosity of $\log (L_{\rm bol}/{\rm erg\,s^{-1}})$\,=\,44.93. High-resolution X-ray spectra revealed the presence of a multi-phase warm absorber with $N_{\rm H}$\,$\sim$\,10$^{22}$--$10^{23}$ cm$^{-2}$, ionisation of $\log\xi$\,$\sim$\,0.5--2.5, and outflow velocities of $\sim$\,1,000--2,000 km s$^{-1}$ \citep[][]{Sanfrutos2016}. The absorbers are likely co-spatial with the inner torus, forming a clumpy multi-phase outflow.

\noindent \textbf{IRAS\,09149--6206} is a nearby ($z$\,=\,0.0573) Seyfert 1 galaxy \citep[][]{Perez1989} hosting a supermassive BH with mass $\log(M_{\rm BH}/M_\odot)$\,=\,8.0$\pm$0.6, and likely accreting close to the Eddington limit \citet[][]{Walton2020}.
\citet{Ricci2017} and \citet[][]{Walton2020} studied the X-ray properties of this AGN and reported on a persistent UTWA with column density up to $N_{\rm H}$\,$\sim$\,6\,$\times10^{22}$ cm$^{-2}$ and ionisation $\log\xi$\,$\sim$\,2--3, showing variability on timescales of months.
\begin{table*}
\centering
\caption{Summary of the properties for the AGN sample displaying UTWAs, ordered by Right Ascension. The first group of sources are those reported in this work, while the second group includes sources from the literature.}
\label{tab:summary_agn_full}
\begin{tabular}{lccccccccc}
\hline
\hline
Name & Optical & $z$ & $\log M_{\rm BH}$ & $\log L_{\rm bol}$ & $\lambda_{\rm Edd}$ & $\log L_{0.5-2\,\rm{keV}}$ & $\log L_{2-10\,\rm{keV}}$ & UFO & Ref. \\
 & Type & & ($M_\odot$) & (erg s$^{-1}$) & & (erg s$^{-1}$) & (erg s$^{-1}$) & & \\
\hline\noalign{\smallskip}
WISEA\,J0232+20&Sy\,1&0.029&7.8&45.2&0.2&42.2/43.7 &42.4/42.9&No &(1)\\[0.5ex]%ok
SDSS\,J0802+31&Sy\,1&0.04& 7.7&44.0&0.02&42.7/43.0&42.8/43.0&No&(1) \\[0.5ex]%ok
SDSS\,J0809+46&QSO&0.655&8.5&46.5&0.9& 44.3/44.4&44.3/44.3&Yes& (1) \\[0.5ex]%ok
PG\,1535+547&Sy\,1&0.038&7.2&44.6&0.2&42.8/44.1&42.4/43.3&Yes&(1)\\[0.5ex]%ok
\hline \noalign{\smallskip}
ESO\,323$-$G77   & Sy\,1.2 & 0.015 & 7.4 & 44.9 & 0.3 & 42.7 & 42.9 & No & (2,3) \\[0.5ex]%ok
NGC\,985         & Sy\,1   & 0.043 & 8.0 & 44.7 & 0.04 & 43.2 &43.6 & No & (4) \\[0.5ex]%okep
1E\,0754.6+3928  & NLS1   & 0.096 & 8.2 & 45.4 & 0.1 & 44.2 & 44.3 & Yes & (5,6) \\[0.5ex]%ok
IRAS\,09149$-$6206 & Sy\,1 & 0.057 & 8.0 & 45.9 & 0.4 & 44.2 & 44.2 & No & (7) \\[0.5ex]%okep
PG\,1114+445     & QSO & 0.144 & 8.8 & 45.7 & 0.07 & 44.4 & 44.2 & Yes & (8) \\[0.5ex]%ok
Mrk\,704         & Sy\,1   & 0.029 & 7.9 & 44.3 & 0.02 & 43.4 & 43.4 & No & (9,10) \\[0.5ex]%okep
3C\,445          & BLRG   & 0.057 & 8.3 & 45.1 & 0.05 & 43.9 & 44.1 & No & (11,12) \\[0.5ex]%ok
Mrk\,304         & Sy\,1   & 0.066 & 8.4 & 44.8 & 0.02 & 43.6 & 43.7 & No & (13,14) \\[0.5ex]%okep[0.5ex]

\hline
\end{tabular}
\flushleft{\small {\bf Notes:} Luminosities are intrinsic (unabsorbed) and expressed in $\log(L/{\rm erg\,s^{-1}})$. $\lambda_{\rm Edd} = L_{\rm bol}/L_{\rm Edd}$. For sources analysed in this work, the reported luminosities correspond to the minimum and maximum values derived with the \texttt{clum} component in \textsc{XSPEC} from the best-fit model. {\bf References:}; (1) this work; (1) \citealt{Wang2007}; (3) \citealt{Jimenez2008}; (4) \citealt{Bianchi2009};  (5) \citealt{Berton2015}; (6) \citealt{Middei2020}; (7)
\citealt{Walton2020}; (8) \citealt{Serafinelli2019}; (9) \citealt{Matt2011}; (10) \citealt{Laha2011}; (11) \citealt{Marchesini2004}; (12) \citealt{Braito2011}; (13) \citealt{Piconcelli2004}; (14) \citealt{Piotrovich2015}.}
\end{table*}

\begin{table*}
\centering
\caption{Properties of the UTWA retrieved from the literature.}
\begin{tabular}{lccc@{\hspace{30pt}}lccc}
\hline
\hline
 Name &$\log N_{\rm H}$ & $\log \xi$ & Ref. & Name &$\log N_{\rm H}$ & $\log \xi$ & Ref. \\
\hline \noalign{\smallskip}
ESO\,323$-$77\,(a)$^*$ & $22.5\pm0.1$ & $2.7\pm0.2$ & Sa16 & 1E\,0754.6+3928\,(1) &$22.85^{+0.05}_{-0.07}$ & $2.00\pm0.05$ & Mi20 \\[0.5ex]
ESO\,323$-$77\,(b)$^*$ & $22.9\pm0.2$ & $1.6\pm0.2$ & Sa16 &1E\,0754.6+3928\,(2) & $22.60^{+0.05}_{-0.06}$ & $1.50\pm0.07$ & Mi20\\[0.5ex]
NGC\,985\,(1a)$^{*\dagger}$ & $22.41^{+0.12}_{-0.16}$ & $1.07^{+0.04}_{-0.11}$ & Eb21& IRAS\,09149$-$6206\,(a)$^\dagger$ & $22.79\pm0.02$ & $2.00^{+0.01}_{-0.02}$ & Wa20 \\ [0.5ex]
NGC\,985\,(1b)$^*$ & $22.30^{+0.27}_{-0.19}$ & $2.44\pm0.08$ & Eb21 & IRAS\,09149$-$6206\,(b) & $22.81^{+0.08}_{-0.10}$ & $3.44^{+0.04}_{-0.06}$ & Wa20 \\ [0.5ex]
NGC\,985\,(2a)$^{*\dagger}$ & $22.89^{+0.09}_{-0.11}$ & $1.68\pm0.05$ & Eb21& PG\,1114+445$^{\dagger}$ & $22.9^{+0.3}_{-0.1}$ & $1.4^{+0.6}_{-0.2}$ & Se21 \\ [0.5ex]
NGC\,985\,(2b)$^*$ & $22.36^{+0.36}_{-0.25}$ & $2.51\pm0.06$ & Eb21& Mrk\,704\,(1)$^{\dagger}$ &$22.94^{+0.09}_{-0.05}$ & $1.99^{+0.06}_{-0.05}$ & Ma11 \\ [0.5ex]
NGC\,985\,(3a)$^{*\dagger}$ & $22.45^{+0.08}_{-0.11}$ & $0.90\pm0.17$ & Eb21& Mrk\,704\,(2)$^{\dagger}$ &$23.12^{+0.06}_{-0.09}$ & $2.09^{+0.18}_{-0.07}$ & Ma11 \\ [0.5ex]
NGC\,985\,(3b)$^*$ & $22.56^{+0.12}_{-0.18}$ & $2.58\pm0.06$ & Eb21& 3C\,445 &$23.28^{+0.01}_{-0.02}$ &  $1.10^{+0.10}_{-0.24}$ & Br11 \\ [0.5ex]
NGC\,985\,(4)$^*$ & $23.0^{+0.2}_{-0.4}$ & $2.93\pm0.02$ & Eb21& Mrk\,304 &$22.95^{+0.02}_{-0.03}$ & $1.95\pm0.06$ & Pi04 \\ [0.5ex]

\hline
\end{tabular}\label{AppT}
\flushleft{\small {\bf Notes:} for each source, progressive numbers refer to UTWAs detected in different epochs, while letters to different UTWA components detected in the same epoch. The dagger $(\dagger)$ superscript indicates that the absorber is partially covering the X-ray source. The asterisk ($*$) flags results obtained through high resolution grating spectra. The properties of the UTWA in PG\,1114+445 are derived from the stacking of {\it Swift} spectra over a period of 22 months. {\bf References:} Sa16 \citep{Sanfrutos2016}; Eb21 \citep{Ebrero2021}; Mi20 \citep{Middei2020}; Wa20 \citep{Walton2020}; Se21 \citep{Serafinelli2021}; Ma11 \citep{Matt2011}; Br11 \citep{Braito2011}; Pi04 \citep{Piconcelli2004}.}
\end{table*}

\noindent \textbf{Mrk\,704} is a polar-scattered Seyfert 1 galaxy at $z$\,=\,0.029 with a BH mass of $\log M_{\rm BH}$\,$\approx$\,7.8 \citep[][]{Afanasiev2019}. \citet[][]{Matt2011} reported on the extreme variability observed in its soft X-ray emission due to a dramatic change in the warm absorber, based on two \textit{XMM-Newton} observations carried out in 2005 and 2008. In particular, they modelled the spectra with two partially-covering absorbing components and found variations in both column density, ionisation state, and covering factor for each component. The component showing the higher ionisation state can be classified as a UTWA, with $N_{\rm H}$\,$\sim$\,$9\times10^{22}$ cm$^{-2}$ and $\log\xi$\,$\sim$\,2 measured in the first observation, which both increased slightly in the later visit, while the covering factor dropped from $\sim$0.85 to $\sim$0.40 between the two epochs. Interestingly, \citet[][]{Matt2011} also presented the results from a set of four {\it Swift} snapshots taken from January 2006 to January 2007 suggesting a variability time scale of a few months or less in the soft X-ray band. However, since the quality of these spectra is limited, only $N_{\rm H}$ and $\xi$ measurements from the two \textit{XMM-Newton} observations are listed in Table A.1.

\noindent \textbf{PG\,1114+445} is a well-known type 1 quasar at $z$\,=\,0.144 \citep[][]{Piconcelli2005}.  In \citet{Shen2011} the authors derive a BH mass of $\log M_{\rm BH}$\,$\sim$\,8.8 and a bolometric luminosity $\log (L_{\rm bol}/{\rm erg\,s^{-1}})$\,=\,45.7, leading to a $\lambda_{\rm Edd}$\,$\sim$\,0.07.  The X-ray spectral properties of this source were extensively investigated by \citet[][]{Serafinelli2019,Serafinelli2021}.
It exhibits a two-zone  warm absorber with column density $N_{\rm H}$ of a few times $10^{22}$ cm$^{-2}$ and $\log\xi$\,$\sim$\,0.3--0.5, plus a high-ionisation UFO with $v_{\rm out}$\,$\sim$\,0.15$c$. Remarkably, for one of the warm-absorption components, \citet[][]{Serafinelli2019} estimated a velocity of $\sim$\,0.13$c$, i.e. similar to that of the UFO. This suggests a scenario in which the UFO entrains and accelerates the ambient material to comparable velocities, thereby producing a multi-phase UFO. Taking advantage of a \textit{Swift} monitoring campaign, \citet[][]{Serafinelli2021} studied the source during an unprecedented low-flux state, distinguished by the long-lasting presence of a UTWA with average $\log(N_{\rm H}/{\rm cm^{-2}})$\,=\,22.9$_{-0.1}^{+0.3}$ and $\log \xi$\,=\,1.4$_{-0.2}^{+0.6}$ (see Table A.1), even though the unabsorbed X-ray luminosity showed no significant variation.

\noindent \textbf{Mrk\,304} (also known as PG\,2214+139) is Seyfert 1 galaxy at $z$\,=\,0.066 that was targeted by {\it XMM-Newton} in 2002 for $\sim$\,30 ks. This observation was analysed by \citet{Piconcelli2004} who derived the 2--10 keV luminosity of the source to be L\,$\sim$\,$4.8\times10^{43}$ erg s$^{-1}$. Moreover, the authors the  presence of complex soft X-ray absorption consisting of two components: the low-ionisation one is a UTWA with $\log(N_{\rm H}/{\rm cm^{-2}})$\,$\sim$\,22.9 and $\log \xi$\,$\sim$\,0.8 (see Table A.1), while the high-ionisation one has a $\log(N_{\rm H}/{\rm cm^{-2}})$\,$\sim$\,22.3 and $\log \xi$\,$\sim$\,2. No other soft X-ray observations exist to investigate the time evolution of these absorbers.

\section{Contours}
\begin{figure*}
   \centering
   \includegraphics[width=\textwidth]{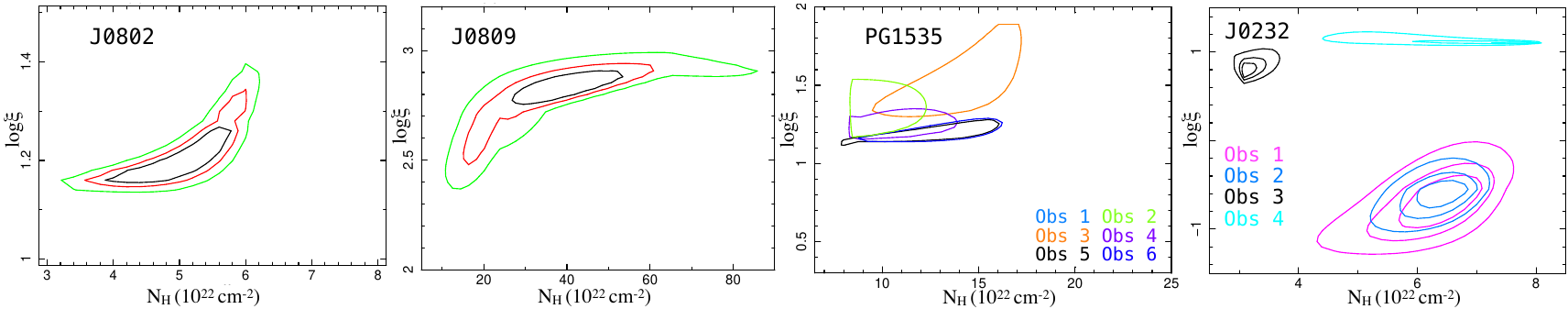}
     \caption{An UTWA has been observed at least in one epoch in all the sources studied in this work. For all the AGN but PG\,1535+547 we report the 68\%, 90\%, and 99\% confidence levels. For the sake of simplicity, for PG\,1535+547 we only show the 90\% confidence regions.}
         \label{contoursall}
\end{figure*}

We show in Figure~\ref{contoursall} the confidence contours of the hydrogen column density and the ionisation parameter obtained from the spectral fits of the four sources analysed in this work.

\end{appendix}
\end{document}